\documentclass{article}
\usepackage{amsfonts}
\usepackage{amsmath} 
\usepackage{url} 
\usepackage{graphicx}

\begin{document}

\title{Polynomials Associated with the Higher Derivatives of the Airy
functions $Ai(z)$ and $Ai^{^{\prime }}(z)$ }
\author{Bernard J. Laurenzi \and Department of Chemistry, UAlbany, The State
University of New York}
\date{September 21, 2011}
\maketitle

\begin{abstract}
The Airy function $Ai(z)$ and its derivative $Ai^{^{\prime }}(z)$ occur in a
large number of applications in Chemistry and Physics. As a result, there is
a continuing interest in the properties of these functions. Recently, there
has been interest in obtaining general expressions for the higher
derivatives of these functions. In this work, general expressions for the
polynomials which are contained in these derivatives are given in terms of
the partial Bell polynomials.
\end{abstract}

In view of the continuing occurrence of the Airy functions $Ai(z)$ and $%
Ai^{^{\prime }}(z)$ in a large number of mathematical and physical
applications, \cite{V} interest in these functions among researchers in
Mathematics, Physics and Chemistry has provided an impetus for the further
exploration of their properties. \ In this paper we investigate the
structural properties of a set of polynomials associated with the
derivatives of the functions $Ai(z)$ and $Ai^{^{\prime }}(z).$

The Airy function defined by the equation%
\[
Ai(z)=\frac{1}{\pi }\int_{0}^{\infty }\cos (\tfrac{1}{3}t^{3}+xt)dt,
\]%
with%
\[
Ai(0)=\frac{1}{3^{2/3}\Gamma (\frac{2}{3})},
\]%
also satisfies the differential equation%
\[
\frac{d^{\,2}Ai(z)}{d\,z^{2}}=z\,Ai(z).
\]%
As a result, direct successive calculations of the higher derivatives of the
Airy function and of its derivative produces the very simple forms 
\begin{eqnarray*}
\frac{d\,^{n}Ai(z)}{dz^{\,n}} &=&\mathcal{P}_{\,n}(z)\,Ai(z)+Q\,_{n}(z)%
\,Ai^{^{\prime }}(z), \\
\frac{d^{\,n}Ai^{^{\prime }}(z)}{dz^{\,n}} &=&\mathcal{P\,}%
_{n+1}(z)\,Ai(z)+Q\,_{n+1}(z)\,Ai^{^{\prime }}(z),
\end{eqnarray*}%
where $\mathcal{P}_{n}(z)$ and $Q\,_{n}(z)$ are polynomials. \ Higher
derivatives of the second linearly independent solution of the Airy
differential equation, i.e. $Bi(z)$ and it derivative $Bi^{\prime }(z)$
produce exactly the same polynomials. \ First studied in detail by
Abramochkin \cite{A}, the Abramochkin polynomials can easily be shown to
satisfy the relations%
\begin{eqnarray*}
\mathcal{P\,}_{n+2}(z) &=&z\,\mathcal{P\,}_{n}(z)+n\,\mathcal{P\,}%
_{n-1}(z)\,, \\
Q\,_{n+2}\left( z\right)  &=&z\,Q\,_{n}\left( z\right) +n\,Q\,_{n-1}(z), \\
\mathcal{P\,}_{n+1}(z) &=&\frac{d\,\mathcal{P\,}_{n}(z)}{d\,z}+z\,Q\,_{n}(z),
\\
Q\,_{n+1}(z) &=&\frac{d\,Q\,_{n}(z)}{d\,z}+\mathcal{P\,}_{n}(z).
\end{eqnarray*}%
The three term recurrence equations shown above do not appear to have been
studied in the past. \ Nor do the usual methods of solving 3$^{\,\underline{%
rd}}$ order difference equations appear to be applicable. \ Generating
functions for these polynomials follow directly from the recurrence
relations and are%
\begin{eqnarray*}
\pi \left[ Bi^{\prime }(z)\,Ai(z+t)-Ai^{\prime }(z)\,Bi(z+t) \right] & = & \sum_{n=0}^{\infty }\frac{t^{n}}{n!}\mathcal{P}_{\,n}(z), \\
\pi \left[ Ai(z)\,Bi(z+t)-Bi(z)\,Ai(z+t) \right] &=&\sum_{n=0}^{\infty }\frac{%
t^{n}}{n!}Q\,_{n}(z).
\end{eqnarray*}%
The last two recurrence equations involving the derivatives of $\mathcal{P}%
_{n}(z)$ and $Q\,_{n}(z)$ are shown in the appendix to be useful in
determining the higher derivatives of the Airy functions. \ The first few
polynomials are given in the Table I. \ As seen in the table, there are
strong regularities in the sequences of these polynomials. \ However, they
are also observed to be complicated functions of the index $n$. \ As a
result one would expect that general expressions for these polynomials will
not be simple. \medskip 

\begin{center}
\begin{tabular}{|c|c|c|}
\hline
$n$ & $\mathcal{P\,}_{n}(z)$ & $Q\,_{n}\left( z\right) $ \\ \hline
$1$ & $0$ & $1$ \\ \hline
$2$ & $z$ & $0$ \\ \hline
$3$ & $1$ & $z$ \\ \hline
$4$ & $z^{2}$ & $2$ \\ \hline
$5$ & $4\,z$ & $z^{2}$ \\ \hline
$6$ & $4+z^{3}$ & $6\,z$ \\ \hline
$7$ & $9\,z^{2}$ & $10+z^{3}$ \\ \hline
$8$ & $28\,z+z^{4}$ & $12\,z^{2}$ \\ \hline
$9$ & $28+16\,z^{3}$ & $52\,z+z^{4}$ \\ \hline
$10$ & $100\,z^{2}+z^{5}$ & $80+20\,z^{3}$ \\ \hline
$11$ & $280\,z+25\,z^{4}$ & $160\,z^{2}+z^{5}$ \\ \hline
$12$ & $280+260\,z^{3}+z^{6}$ & $600\,z+30\,z^{4}$ \\ \hline
$13$ & $1380\,z^{2}+36\,z^{5}$ & $880+380\,z^{3}+z^{6}$ \\ \hline
$14$ & $3640\,z+560\,z^{4}+z^{7}$ & $2520\,z^{2}+4228\,z^{5}$ \\ \hline
$15$ & $3640+4760z^{3}+49z^{6}$ & $8680z+770z^{4}+z^{7}$ \\ \hline
$16$ & $22960z^{2}+1064z^{5}+z^{8}$ & $12320+7840z^{3}+56z^{6}$ \\ \hline
$17$ & $58240z+13160z^{4}+64z^{7}$ & $46480z^{2}+1400z^{5}+z^{8}$ \\ \hline
$18$ & $58240+99120z^{3}+1848z^{6}+z^{9}$ & $151200z+20160z^{4}+72z^{7}$ \\ 
\hline
\end{tabular}%
\bigskip\ 

Table I\medskip
\end{center}

Indeed, attempts to obtain general expressions for these polynomials have
proven to be challenging. \ In computing the $n$ $^{\text{\underline{th}}}$
derivatives of the Airy functions it is useful to note that they are related
to the modified Bessel functions of the second kind $K_{\nu }(\zeta )$ of
fractional order $\nu $, (sometimes called Basset functions) by the
relations \cite{AandS}\smallskip \medskip 
\begin{eqnarray*}
Ai(z) &=&\frac{1}{\pi }\sqrt{\frac{z}{3}}K_{1/3}(\zeta ), \\
Ai^{^{\prime }}(z) &=&-\frac{1}{\pi }\frac{z}{\sqrt{3}}K_{2/3}(\zeta ), \\
\zeta &=&\tfrac{2}{3}z^{\,3/2}.
\end{eqnarray*}%
In the computation of the $\mathcal{P}_{n}(z)$ and $Q\,_{n}(z)$ polynomials
it is sufficient to consider the $n$ $^{\text{\underline{th}}}$ derivative
of the function $Ai^{^{\prime }}(z).$ \ In terms of the Bessel function $%
K_{2/3}(\zeta )$ these derivatives are given by%
\begin{equation}
\frac{d^{\,n}Ai^{^{\prime }}(z)}{d\,z\,^{n}}=-\,\frac{1}{\pi \sqrt{3}}\left[
z\frac{d^{\,n}K_{2/3}(\zeta )}{d\,z^{\,n}}+n\frac{d^{\,n-1}K_{2/3}(\zeta )}{%
d\,z\,^{n-1}}\right] .  \label{eq1}
\end{equation}

The $n$ $^{\text{\underline{th}}}$ derivative of a function $f\,(\zeta (z))$
which is an implicit function of the variable $z$ can be given by Fa\`{a} di
Bruno's formula \cite{Faa}

\begin{equation}
\frac{d^{\,n}f\,(\zeta (z))}{d\,z\,^{n}}=\sum_{k=0}^{n}\frac{%
d^{\,k}f\,(\zeta )}{d\zeta ^{\,k}}\,B_{n,\,k}(\frac{d\zeta }{d\,z},\frac{%
d^{\,2}\zeta }{d\,z^{2}},\cdots ,\frac{d^{\,n-k+1}\zeta }{d\,z^{n-k+1}}),
\label{eq2}
\end{equation}%
written in terms of the \textit{partial Bell polynomials} $B_{n,\,k}$ \cite%
{Bell} where the polynomials $B_{\nu ,\,k}$ (of total weight $\nu $ and
degree $k$ in the variables $\ x_{1},x_{2},\cdots ,x_{n-k+1}\,$) are given by%
\[
B_{\nu ,\,k}(x_{1},x_{2},\cdots \,x_{\nu -k+1})=\sum \frac{\nu !}{%
k_{1}!k_{2}!\cdots k_{l}!}\left( \frac{x_{1}}{1!}\right) ^{k_{1}}\left( 
\frac{x_{2}}{2!}\right) ^{k_{2}}\cdots \left( \frac{x_{l}}{l!}\right)
^{k_{l}}, 
\]%
and the sum ranges over all possible values of the integers $k_{i}$ subject
to the conditions 
\begin{eqnarray*}
\nu -k+1 &=&l, \\
k_{1}+k_{2}+\cdots +k_{l} &=&k, \\
k_{1}+2\,k_{2}+\cdots +l\,k_{l} &=&\nu .
\end{eqnarray*}%
The partial Bell polynomials can easily be obtained using the generating
function%
\begin{equation}
\sum_{\nu =k}^{\infty }B_{\nu ,\,k}(x_{1},x_{2},\cdots ,x_{\nu -k+1})\frac{%
t^{\nu }}{\nu !}=\frac{1}{k!}\left( \sum_{\mu =1}^{\nu -k+1}\frac{x_{\mu
}t^{\mu }}{\mu !}\right) ^{k},  \label{eq3}
\end{equation}%
from which one gets 
\[
B_{n,\,k}(x_{1},x_{2},\cdots ,x_{n-k+1})=n!\cdot ([t^{n}]\;\frac{1}{k!}%
\left( \sum_{\mu =1}^{n-k+1}\frac{x_{\mu }\,t^{\mu }}{\mu !}\right) ^{k}), 
\]%
where the notation $([t^{n}]\;\frac{1}{k!}\left( \sum_{\mu =1}^{n-k+1}\frac{%
x_{\mu }t^{\mu }}{\mu !}\right) ^{k})$ stands for the coefficient of $t^{n}$
in the sum on the right hand side of the generating function's relation Eq.
(3). \ It is useful to note the special cases of $B_{n,\,k}$ shown in the
Table II. These values have been obtain from relations due to Cvijovic \cite%
{C}.

\begin{center}
\begin{tabular}{|c|c|}
\hline
$k$ & $B_{n,\,k}$ \\ \hline
$0$ & $0$ \\ \hline
$1$ & $x_{n}$ \\ \hline
$n$ & $x_{1}^{n}$ \\ \hline
$n-1$ & $\tbinom{m}{2}x_{1}^{n-2}x_{2}$ \\ \hline
$n-2$ & $\frac{3}{4}\tbinom{m}{3}x_{1}^{n-3}[(n-3)x_{2}^{2}/x_{1}+\frac{4}{3}%
x_{3}]$ \\ \hline
\end{tabular}
\ \medskip

Table II\medskip
\end{center}

Alternately, the Bell polynomials can be computed using software packages
such as \textit{Maple} (version 15) or \textit{Mathematica} which contain
the partial Bell polynomials as part of their standard sets of mathematical
functions.

Using Eq. (2) we obtain the required derivatives as 
\[
\frac{d^{\,n}Ai^{^{\prime }}(z)}{dz^{\,n}}=\frac{(-1)^{n+1}}{\pi \sqrt{3}%
\;2^{n}\,\left( \frac{3}{2}\zeta \right) ^{\frac{2}{3}(n-1)}}{\small \cdot }%
{\Huge [}\sum_{k=0}^{n}(3\zeta )^{k}\frac{d^{\,k}K_{2/3}(\zeta )}{d\zeta
^{\,k}}B_{n,\,k}(-3!!,-1!!,\cdots ,(2n-2k-3)!!) 
\]%
\[
-2n\,\sum_{k=0}^{n-1}(3\zeta )^{k}\frac{d^{\,k}K_{2/3}(\zeta )}{d\zeta
\,^{k}}B_{n-1,\,k}(-3!!,-1!!,\cdots ,(2n-2k-5)!!){\Huge ]}\ \text{ \ for }%
n\geq 1, 
\]%
where the $\kappa ^{\text{\underline{th}}}$ derivative of $\zeta $ with
respect to $z$ as given by%
\[
\frac{d^{\,\kappa }\zeta }{d\,z^{\,\kappa }}=2(-2)^{-\kappa }(2\kappa -5)!!(%
\tfrac{3}{2}\zeta )^{1-\frac{2}{3}\kappa }, 
\]%
has been used to rescale and simplify the required Bell polynomials (the
double factorial function has values $(2\kappa -5)!!=1$%
$\cdot$%
$3$%
$\cdot$%
$5\cdots (2\kappa -5)$ with $(-3)!!=-1,$and $(-1)!!=1$).

The sums above can be combined if the last term in the first sum is written
out explicitly and use is made of the relation $B_{n,\,n}((-3)!!)=(-1)^{n}.$
\ Written more compactly we have%
\begin{equation}
\frac{d^{\,n}Ai^{^{\prime }}(z)}{d\,z^{\,n}}=\frac{(-1)^{n+1}}{\pi \sqrt{3}%
\,2^{n}\,\left( \frac{3}{2}\zeta \right) ^{\frac{2}{3}(n-1)}}\left[ (-3\zeta
)^{n}\frac{d^{\,n}K_{2/3}(\zeta )}{d\zeta ^{\,n}}+\sum_{k=0}^{n-1}(3\zeta
)^{k}\frac{d^{\,k}K_{2/3}(\zeta )}{d\zeta \,^{k}}\Delta B_{n,\,k}\right] ,
\label{eq4}
\end{equation}%
where we have defined the term $\Delta B_{n,\,k}$ as 
\[
\Delta B_{n,\,k}=B_{n,\,k}(\left( -3\right) !!,\left( -1\right) !!,\cdots
,(2n-2k-3)!!)-2n{\small \cdot }B_{n-1,\,k}(\left( -3\right) !!,\left(
-1\right) !!,\cdots ,(2n-2k-5)!!), 
\]%
since $x_{i}=(2i-5)!!$. \ The derivatives of the modified Bessel function in
Eq. (4) can be obtained from the well known relation \cite{Bess} for the $n$ 
$^{\text{\underline{th}}}$ derivative of $K_{\nu }(\zeta )$ i.e.%
\begin{equation}
\frac{d^{\,n}K_{\nu }(\zeta )}{d\zeta ^{n}}=\tfrac{\left( -1\right) ^{n}}{%
2^{n}}\sum_{i=0}^{n}\tbinom{n}{i}\,K_{\nu +2i-n}(\zeta ).  \label{eq5}
\end{equation}%
We note that Bessel functions of order higher and lower than $\nu $ occur in
this sum. \ With $\nu =2/3$ all of these terms can be reduced to ones which
contain the functions $K_{1/3}(\zeta )$ and $K_{2/3}(\zeta )$ by means of
the relation

\begin{eqnarray}
K_{\eta }(\zeta ) &=&K_{\eta +2\mu \;}(\zeta )\;\left[ \sum_{p=0}^{\mu -1}%
\tbinom{\mu -1+p}{2p}\tfrac{\Gamma (\eta +\mu +p)}{\Gamma (\eta +\mu -p)}%
 {2}{\zeta }^{2\,p}\right]  \label{eq6} \\
&&-K_{\eta +2\mu -1\;}(\zeta )\;\left[ \sum_{p=0}^{\mu -1}\tbinom{\mu +p}{%
2p+1}\tfrac{\Gamma (\eta +\mu +1+p)}{\Gamma (\eta +\mu -p)}
{2}{\zeta }^{2\,p+1}\right] ,\text{ \ \ for \ \ }\mu >0.  \nonumber
\end{eqnarray}%
The equation above allows Bessel functions with negative order to be
replaced with those with positive order. \ Equation (6) has been obtained by
repeated application of the recurrence relation \cite{Bess} for $K_{\nu
}(\zeta )$ i.e.

\[
K_{\nu }(\zeta )=K_{\nu +2}(\zeta )-\left( \tfrac{2}{\zeta }\right) (\nu
+1)K_{\nu +1}(\zeta ). 
\]%
Using Eq. (5) for the derivatives we have 
\begin{equation}
\frac{d^{\,n}Ai^{^{\prime }}(z)}{d\,z\,^{n}}=\frac{(-1)^{n+1}}{\pi \sqrt{3}%
\,2^{n}\,\left( \frac{3}{2}\zeta \right) ^{\frac{2}{3}(n-1)}}\left[ (-\tfrac{%
3}{2}\zeta )^{n}\sum_{i=0}^{n}\tbinom{n}{i}K_{2/3+2i-n}(\zeta
)+\sum_{k=0}^{n-1}(-\tfrac{3}{2}\zeta )^{k}\,\Delta B_{n,\,k}\sum_{i=0}^{k}%
\tbinom{k}{i}K_{2/3+2i-k}(\zeta )\right] .  \label{eq7}
\end{equation}

\subsection{Calculation of the Bell-free terms of equation (7)}

In order to ease the calculation of the sums in Eq. (7) and without loss of
generality (cf. the appendix) we consider only even values of $n.$ With $%
n=2m,$ for $m\geq 1$, the resulting calculation of $d^{\,2m}Ai^{^{\prime
}}(z)/dz^{\,2m}\ $eventually yields $\mathcal{P}_{2m+1}(z)$ and $%
Q\,_{2m+1}\left( z\right) .$

Under these circumstances the Bell-free term in Eq. (7) which corresponds to
the even order derivatives of $K_{2/3}(\zeta )$ is given by%
\begin{equation}
\sum_{i=0}^{2m}\tbinom{2m}{i}K_{2/3+2i-2m}(\zeta )=\tbinom{2m}{m}%
K_{2/3}(\zeta )+\sum_{i=0}^{m-1}\tbinom{2m}{i}K_{2/3+2\,i-2m}(\zeta
)+\sum_{i=m+1}^{2m}\tbinom{2m}{i}K_{2/3+2\,i-2m}(\zeta ).  \label{eq8}
\end{equation}%
In the first sum of Eq. (8), the orders of the Bessel functions $2/3+2\,i-2m$
within the range $0\leq i\leq m-1$ are negative and every Bessel function
therein can be replaced with a term containing the functions $K_{1/3}(\zeta
) $ and $K_{2/3}(\zeta ).$ \ Choosing $\nu =2/3+2\,i-2m,$ and \ $\nu +2\mu
=2/3 $ in Eq. (6) requires that $\mu =m-i$. \ As a result we get 
\begin{eqnarray}
K_{2/3+2\,i-2m}(\zeta ) &=&-\,K_{1/3}(\zeta )\sum_{p=0}^{m-i-1}\tbinom{m-i+p%
}{2p+1}\tfrac{\Gamma (5/3-m+i+p)}{\Gamma (2/3-m+i-p)}{2}{\zeta
}^{2p+1}  \label{eq9} \\
&&+K_{2/3}(\zeta )\sum_{p=0}^{m-i-1}\tbinom{m-i-1+p}{2p}\tfrac{\Gamma
(2/3-m+i+p)}{\Gamma (2/3-m+i-p)}{2}{\zeta }^{2p},  \nonumber
\end{eqnarray}%
where we have used Eq. (5) and the \textit{parity} property of the Bessel
functions i.e. $K_{-\nu }(\zeta )=K_{\nu }(\zeta )$ for the $K_{1/3}(\zeta )$
function. \ \smallskip

In the second sum in Eq. (8) the orders $2/3+2\,i-2m$ within the range $%
m+1\leq i\leq 2m$ are positive$.$ Using the parity property of the Bessel
functions of order $\nu $ we get 
\[
K_{2/3+2\,i-2m}(\zeta )=K_{-2/3-2\,i+2m}(\zeta ). 
\]%
In this case we set $\nu =-2/3-2\,i+2m,$ and \ $\nu +2\mu =-2/3$ which gives 
$\mu =i-m$ and using Eq. (6) we raise these indexes to get 
\begin{eqnarray*}
K_{\,-2/3-2\,i+2m}(\zeta ) &=&K_{\,-2/3}(\zeta )\sum_{p=0}^{i-m-1}\tbinom{%
i-m-1+p}{2p}\tfrac{\Gamma (-2/3-i+m+p)}{\Gamma (-2/3-i+m-p)}
{2}{\zeta }^{2p} \\
&&-K_{\,-5/3}(\zeta )\sum_{p=0}^{i-m-1}\tbinom{i-m+p}{2p+1}\tfrac{\Gamma
(1/3-i+m+p)}{\Gamma (-2/3-i+m-p)}{2}{\zeta }^{2p+1}.
\end{eqnarray*}%
Employing the relations $K_{\,-2/3}(\zeta )=K_{\,2/3}(\zeta )$ and $%
K_{\,-5/3}(\zeta )=K_{1/3}(\zeta )+(\frac{2}{3})(\frac{2}{\zeta }%
)\,K_{\,2/3}(\zeta ),$ the latter expression can be written as%
\[
K_{\,-2/3-2\,i+2m}(\zeta )=-K_{1/3}(\zeta )\sum_{p=0}^{i-m-1}\tbinom{i-m+p}{%
2p+1}\tfrac{\Gamma (1/3-i+m+p)}{\Gamma (-2/3-i+m-p)}{2}{\zeta
}^{2p+1} 
\]%
\begin{equation}
+K_{\,2/3}(\zeta )\sum_{p=0}^{i-m-1}\left[ \tbinom{i-m-1+p}{2p}\tfrac{%
\Gamma (-2/3-i+m+p)}{\Gamma (-2/3-i+m-p)}{2}{\zeta }^{2p}-(%
\tfrac{2}{3}){2}{\zeta }^{2}\sum_{p=0}^{i-m-1}\tbinom{i-m+p}{%
2p+1}\tfrac{\Gamma (1/3-i+m+p)}{\Gamma (-2/3-i+m-p)}\right] 
{2}{\zeta }^{2p}.  \label{eq10}
\end{equation}%
Combining Eqs. (9, 10) and reversing the order of the resulting double sums,
Eq. (8) can be written in the more compact form \ 
\[
(\tfrac{3\zeta }{2})^{2m}\sum_{i=0}^{2m}\tbinom{2m}{i}K_{2/3+2i-2m}(\zeta
)=-K_{1/3}(\zeta )\,3^{2m}\,\,\sum_{p=0}^{m-1}C_{1}^{(0)}(m,p)\left( \tfrac{%
\zeta }{2}\right) ^{2m-2p-1} 
\]%
\[
+K_{2/3}(\zeta )\left[ \tbinom{2m}{m}\left( \tfrac{3\zeta }{2}\right)
^{2m}+\,3^{2m}\sum_{p=0}^{m-1}C_{2}(m,p;\zeta )\left( \tfrac{\zeta }{2}%
\right) ^{2m-2p}\right] , 
\]%
or reindexing the powers of $\zeta $ in these sums we have 
\[
(\tfrac{3}{2}\zeta )^{2m}\sum_{i=0}^{2m}\tbinom{2m}{i}K_{2/3+2i-2m}(\zeta
)=-\left( \tfrac{2}{\zeta }\right) K_{1/3}(\zeta
)\,3^{2m}\sum_{q=1}^{m}C_{1}^{(0)}(m,m-q)\left( \tfrac{\zeta }{2}\right)
^{2q} 
\]%
\begin{equation}
+K_{2/3}(\zeta )\left[ \tbinom{2m}{m}\left( \tfrac{3\zeta }{2}\right)
^{2m}+\,3^{2m}\sum_{q=1}^{m}C_{2}(m,m-q;\zeta )\left( \tfrac{\zeta }{2}%
\right) ^{2q}\right] ,  \label{eq11}
\end{equation}%
where the coefficient $C_{1}^{(0)}(m,p)$ and the quantity $C_{2}(m,p;\zeta )$
are given by the finite sums 
\[
C_{1}^{(0)}(m,p)=\sum_{i=0}^{m-1-p}\left[ \tbinom{2m}{i}\tbinom{m+p-i}{2p+1}%
\tfrac{\Gamma (5/3-m+p+i)}{\Gamma (2/3-m-p+i)}-\tbinom{2m}{m+p+1+i}\tbinom{%
i+2p+1}{2p+1}\tfrac{\Gamma (8/3+2p+i)}{\Gamma (5/3+i)}\right] , 
\]%
and%
\[
C_{2}(m,p;\zeta )=\sum_{i=0}^{m-1-p}{\LARGE [}\tbinom{2m}{i}\tbinom{m-1+p-i%
}{2p}\tfrac{\Gamma (2/3-m+p+i)}{\Gamma (2/3-m-p+i)} 
\]%
\[
+\tbinom{2m}{m+p+1+i}\tbinom{i+2p+1}{2p+1}\tfrac{\Gamma (8/3+2p+i)}{\Gamma
(5/3+i)}\{\tfrac{(2p+1)}{\left( 2p+1+i\right) (5/3+i)}+\tfrac{2}{3}\left( 
\tfrac{2}{\zeta }\right) ^{2}\}{\LARGE ]}. 
\]

\subsection{Calculation of the Bell-containing terms of equation (7)}

In this case, both even and odd order derivatives of the Bessel function $%
K_{2/3}(\zeta )$ occur. \ Consequently, when the left hand side of the
equation below has been rewritten as sums corresponding to the even and odd
values of $k$, the following expression results 
\[
\sum_{k=0}^{2m-1}(-\tfrac{3}{2}\zeta )^{k}\,\Delta B_{2m,\,k}\sum_{i=0}^{k}%
\tbinom{k}{i}K_{2/3+2i-k}(\zeta )=\sum_{j=0}^{m-1}\left( \tfrac{3\zeta }{2}%
\right) ^{2\,j}\Delta B_{2m,\,2\,j}\sum_{i=0}^{2\,j}\tbinom{2\ j}{i}%
K_{2/3+2\,i-2\,j}(\zeta ) 
\]%
\begin{equation}
-\sum_{j=0}^{m-1}\left( \tfrac{3\zeta }{2}\right) ^{2\,j+1}\Delta
B_{2m,\,2\,j+1}\sum_{i=0}^{2\,j+1}\tbinom{2\,j+1}{i}K_{-1/3+2\,i-2\
j}(\zeta ).  \label{eq12}
\end{equation}%
We note that the first sum on the right hand side of Eq. (8) contains a Bell
term with even degree $2j$ and the second sum contains a Bell term with odd
degree $2j+1.$ \ In the relation above it will be convenient to treat the
cases with $j=0$ separately where $j$ has the range $0$ $\leq j\leq m-1$. \
As a result we rewrite Eq. (12) as%
\[
\sum_{k=0}^{2m-1}(-\tfrac{3}{2}\zeta )^{k}\,\Delta B_{2m,\,k}\sum_{i=0}^{k}%
\tbinom{k}{i}K_{2/3+2i-k}(\zeta )= 
\]%
\[
\Delta B_{2m,\,0}\,K_{2/3}(\zeta )-\left( \tfrac{3\zeta }{2}\right) \Delta
B_{2m,\,1}\left[ 2K_{1/3}(\zeta )+\tfrac{2}{3}\left( \tfrac{2}{\zeta }%
\right) K_{2/3}(\zeta )\right] 
\]%
\[
+H(m-2){\Huge \{}\sum_{j=1}^{m-1}\left( \tfrac{3\zeta }{2}\right)
^{2\,j}\Delta B_{2m,\,2\,j}\sum_{i=0}^{2\,j}\tbinom{2\ j}{i}%
K_{2/3+2\,i-2\,j}(\zeta ) 
\]%
\[
-\sum_{j=1}^{m-1}\left( \tfrac{3\zeta }{2}\right) ^{2\,j+1}\Delta
B_{2m,\,2\,j+1}\sum_{i=0}^{2\,j+1}\tbinom{2\,j+1}{i}K_{-1/3+2\,i-2\
j}(\zeta ){\Huge \}}, 
\]%
where $H$ is the Heaviside \cite{Heavi} step function has been inserted to
ensure in the case where $m=1$ that the double sums don't contribute to the
left hand side of the equation above. \ Using the values of $\Delta
B_{2m,\,0}=0$ and $\Delta B_{2m,\,1}=-5(4m-7)!!$ we get%
\[
\sum_{k=0}^{2m-1}(-\tfrac{3}{2}\zeta )^{k}\,\Delta B_{2m,\,k}\sum_{i=0}^{k}%
\tbinom{k}{i}K_{2/3+2i-k}(\zeta )= 
\]%
\[
10(4m-7)!!\left( \tfrac{3\zeta }{2}\right) \left[ K_{1/3}(\zeta )+\left( 
\tfrac{2}{3\zeta }\right) K_{2/3}(\zeta )\right] 
\]%
\[
+H(m-2)\left\{ 
\begin{array}{c}
\sum_{j=1}^{m-1}\left( \tfrac{3\zeta }{2}\right) ^{2\,j}\Delta
B_{2m,\,2\,j}\sum_{i=0}^{2\,j}\tbinom{2\ j}{i}K_{2/3-2\,j+2\,i}(\zeta ) \\ 
-\sum_{j=1}^{m-1}\left( \tfrac{3\zeta }{2}\right) ^{2\,j+1}\Delta
B_{2m,\,2\,j+1}\sum_{i=0}^{2\,j+1}\tbinom{2\,j+1}{i}K_{-1/3-2\
j+2\,i}(\zeta )%
\end{array}%
\right\} . 
\]

The next task being the reduction of the $K$ functions with orders $%
2/3-2j+2i $ (in the Bell sum with even degree) and $-1/3-2j+2i$ (in the Bell
sum with odd degree) to orders $1/3$ and $2/3$ .

\subsubsection{Analysis of the Bell sum of equation (12) with even degree}

We start with the sum of Bessel functions contained in the Bell sum of even
degree of Eq. (12) i.e. 
\begin{equation}
\sum_{i=0}^{2\,j}\tbinom{2\,j}{i}K_{2/3+2\,i-2\,j}(\zeta )=\sum_{i=0}^{j-1}%
\tbinom{2\,j}{i}K_{2/3+2\,i-2\,j}(\zeta )+\sum_{i=j+1}^{2\,j}\tbinom{2\,j}{i%
}K_{2/3+2\,i-2\,j}(\zeta )  \label{eq13}
\end{equation}%
\[
+\tbinom{2\,j}{j}K_{2/3}(\zeta ),\text{ \ \ \ for\ \ \ \ }j\geq 1, 
\]%
where we have separated the sum of $K$ functions in Eq. (13) into those with
positive order greater than $2/3,$ negative order less than $2/3$ and order $%
2/3$ respectively. \ As seen above in connection with Eq. (6) we have for $%
K_{2/3-2\,j+2\,i}(\zeta )$ in the first sum of Eq. (13) 
\[
K_{2/3-2\,j+2\,i}(\zeta )=-K_{1/3}(\zeta )\sum_{p=0}^{j-i-1}\tbinom{j-i+p}{%
2p+1}\tfrac{\Gamma (5/3-j+i+p)}{\Gamma (2/3-j+i-p)}\left( \tfrac{2}{\zeta }%
\right) ^{2p+1} 
\]%
\begin{equation}
+K_{2/3}(\zeta )\sum_{p=0}^{j-i-1}\tbinom{j-i-1+p}{2p}\tfrac{\Gamma
(2/3-j+i+p)}{\Gamma (2/3-j+i-p)}\left( \tfrac{2}{\zeta }\right) ^{2p}.
\label{eq14}
\end{equation}%
In the second sum in Eq. (13) we have replaced order $2/3-2j+2i$ with $%
-2/3+2j-2i.$ We get for $K_{-2/3+2\,j-2\,i}(\zeta )$

\bigskip 
\[
K_{-2/3+2\,j-2\,i}(\zeta )=K_{2/3}(\zeta )\sum_{p=0}^{j-i-1}\tbinom{i-j-1+p%
}{2p}\tfrac{\Gamma (-2/3-i+j+p)}{\Gamma (-2/3-i+j-p)}\left( \tfrac{2}{\zeta }%
\right) ^{2p} 
\]%
\[
-K_{-5/3}(\zeta )\sum_{p=0}^{i-j-1}\tbinom{i-j+p}{2p+1}\tfrac{\Gamma
(1/3-i+j+p)}{\Gamma (-2/3-i+j-p)}\left( \tfrac{2}{\zeta }\right) ^{2p+1}. 
\]%
Using the relation $K_{-5/3}(\zeta )=K_{1/3}(\zeta )+(2/3)(2/\zeta
)K_{2/3}(\zeta )$ the last expression becomes 
\[
K_{-2/3+2\,j-2\,i}(\zeta )=-K_{1/3}(\zeta )\sum_{p=0}^{i-j-1}\tbinom{i-j+p}{%
2p+1}\tfrac{\Gamma (1/3-i+j+p)}{\Gamma (-2/3-i+j-p)}\left( \tfrac{2}{\zeta }%
\right) ^{2p+1} 
\]%
\begin{equation}
+K_{2/3}(\zeta )\sum_{p=0}^{i-j-1}[\tbinom{i-j-1+p}{2p}\tfrac{\Gamma
(-2/3-i+j+p)}{\Gamma (-2/3-i+j-p)}-\tfrac{2}{3}\left( \tfrac{2}{\zeta }%
\right) ^{2}\tbinom{i-j+p}{2p+1}\tfrac{\Gamma (1/3-i+j+p)}{\Gamma
(-2/3-i+j-p)}]\left( \tfrac{2}{\zeta }\right) ^{2p}.  \label{eq15}
\end{equation}%
Using Eq. (14) and Eq. (15) we get after reversing the order of summation in
Eq. (13) a relation in complete analogy to the result found in connection
with the Bell free term in Eq. (11) i.e.

\begin{equation}
\sum_{i=0}^{2j}\tbinom{2\,j}{i}K_{2/3-2j+2i}(\zeta )=-\,\left( \tfrac{2}{%
\zeta }\right) K_{1/3}(\zeta )\sum_{p=0}^{j-1}C_{1}^{(0)}(j,p)\left( \tfrac{%
2}{\zeta }\right) ^{2p}+K_{2/3}(\zeta )[\tbinom{2\,j}{j}%
+\sum_{p=0}^{j-1}C_{2}(j,p;\zeta )\left( \tfrac{2}{\zeta }\right) ^{2p}].
\label{eq16}
\end{equation}%
\bigskip As a result, the Bell term of even degree in Eq. (12) can be
written as%
\[
\sum_{j=1}^{m-1}\left( \tfrac{3\zeta }{2}\right) ^{2\,j}\Delta
B_{2m,\,2\,j}\sum_{i=0}^{2\,j}\tbinom{2\ j}{i}K_{2/3-2\,j+2\,i}(\zeta )= 
\]%
\[
-(\tfrac{2}{\zeta })K_{1/3}(\zeta )\sum_{q=1}^{m-1}\left(
\sum_{j=q}^{m-1}3^{2j}\,\Delta B_{2m,\,2j}\,\cdot C_{1}^{(0)}(j,j-q)\right)
\left( \tfrac{\zeta }{2}\right) ^{2q} 
\]%
\begin{equation}
+K_{2/3}(\zeta )\left\{ \sum_{q=1}^{m-1}\left(
\sum_{j=q}^{m-1}3^{2j}\,\Delta B_{2m,\,2j}\,\cdot C_{2}(j,j-q;\zeta
)\right) \left( \tfrac{\zeta }{2}\right) ^{2q}+\sum_{j=1}^{m-1}\tbinom{2j}{j%
}\,\Delta B_{2m,2j}\,\left( \tfrac{3\,\zeta }{2}\right) ^{2j}\right\} .
\label{eq17}
\end{equation}

\subsubsection{Analysis of the Bell sum of equation (12) with odd degree}

We now consider the sum of Bessel functions contained in the Bell sum with
odd degree of Eq. (12). \ As before, we have separated the sum into Bessel
functions with order $1/3,$ and those with positive order greater than and
less than $1/3$ i.e. 
\[
\sum_{i=0}^{2\,j+1}\tbinom{2\,j+1}{i}K_{-1/3-2\,\,j+2\,i}(\zeta
)=\sum_{i=0}^{j-1}\tbinom{2\,j+1}{i}K_{-1/3-2\,j+2\,i}(\zeta
)+\sum_{i=j+1}^{2\,j+1}\tbinom{2\,j+1}{i}K_{1/3+2\,j-2\,i}(\zeta ) 
\]%
\[
+\tbinom{2\,j+1}{j}K_{1/3}(\zeta ),\;\text{\ for\ \ }\;j>0, 
\]%
Using Eq. (6) we obtain for $j>0$,%
\[
K_{-1/3-2\,j+2\,i}(\zeta )=K_{1/3}(\zeta )\sum_{p=0}^{j-1-i}{\LARGE [}%
\tbinom{j-1-i+p}{2p}\tfrac{\Gamma (-1/3-j+i+p)}{\Gamma (-1/3-j+i-p)}-\tfrac{1%
}{3}\left( \tfrac{2}{\zeta }\right) ^{2}\tbinom{j-i+p}{2p+1}\tfrac{\Gamma
(2/3-j+i+p)}{\Gamma (-1/3-j+i-p)}{\LARGE ]}\left( \tfrac{2}{\zeta }\right)
^{2p} 
\]%
\[
-\left( \tfrac{2}{\zeta }\right) K_{2/3}(\zeta )\sum_{p=0}^{j-1-i}\tbinom{%
j-i+p}{2p+1}\tfrac{\Gamma (2/3-j+i+p)}{\Gamma (-1/3-j+i-p)}\left( \tfrac{2}{%
\zeta }\right) ^{2p}, 
\]%
\ \ \ \ \ \ \ \ \ \ \ \ \ \ \ \ \ \ \ \ \ \ \ \ \ \ \ \ \ \ \ \ \ \ \ \ \ \
\ \ \ \ \ \ \ \ \ \ \ \ \ \ \ \ \ \ \ \ \ \ \ \ \ \ \ \ \ \ \ \ \ \ \ \ \ \
\ \ \ \ \ \ \ \ \ \ \ \ \ \ \ \ \ \ \ \ \ \ \ \ \ \ \ \ \ \ \ \ \ \ \ \ \ \
\ \ \ \ \ \ \ \ \ \ \ \ \ \ \ \ \ \ \ \ \ \ \ \ \ \ \ \ \ \ \ \ \ \ \ \ \ \
\ \ \ \ \ \ \ \ \ \ \ \ \ \ \ \ \ \ \ \ \ \ \ \ \ \ \ \ \ \ \ \ \ \ \ \ \ \
\ \ \ \ \ \ \ \ \ \ \ \ \ \ \ \ \ \ \ \ \ \ \ \ \ \ \ \ \ \ \ \ \ \ \ \ \ \
\ \ \ \ \ \ \ \ \ \ \ \ \ \ \ \ \ \ \ \ \ \ \ \ \ \ \ \ \ \ \ \ \ \ \ \ \ \
\ \ \ \ \ \ \ \ \ \ \ \ \ \ \ \ \ \ \ \ \ \ \ \ \ \ \ \ \ \ \ \ \ \ \ \ \ \
\ \ \ \ \ \ \ \ \ \ \ \ \ \ \ \ \ \ \ \ \ \ \ \ \ \ \ \ \ \ \ \ \ \ \ \ \ \
\ \ \ \ \ \ \ \ \ \ \ \ \ \ \ \ \ \ \ \ \ \ \ \ \ \ \ \ \ \ \ \ \ \ \ \ \ \
\ \ \ \ \ \ \ \ \ \ \ \ \ \ \ \ \ \ \ \ \ \ \ \ \ \ \ \ \ \ \ \ \ \ \ \ \ \
\ \ \ \ \ \ \ \ \ \ \ \ \ \ \ \ \ \ \ \ \ \ \ \ \ \ \ \ \ \ \ \ \ \ \ \ \ \
\ \ \ \ \ \ \ \ \ \ \ \ \ \ \ \ \ \ \ \ \ \ \ \ \ \ \ \ \ \ \ \ \ \ \ \ \ \
\ \ \ \ \ \ \ \ \ \ \ \ \ \ \ \ \ \ \ \ \ \ \ \ \ \ \ \ \ \ \ \ \ \ \ \ \ \
\ \ \ \ \ \ \ \ \ \ \ \ \ \ \ \ \ \ \ \ \ \ \ \ \ \ \ \ \ \ \ \ \ \ \ \ \ \
\ \ \ \ \ \ \ \ \ \ \ \ \ \ \ \ \ \ \ \ \ \ \ \ \ \ \ \ \ \ \ \ \ \ \ \ \ \
\ \ \ \ \ \ \ \ \ \ \ \ \ \ \ \ \ \ \ \ \ \ \ \ \ \ \ \ \ \ \ \ \ \ \ \ \ \
\ \ \ \ \ \ \ \ \ \ \ \ \ \ \ \ \ \ \ \ \ \ \ \ \ \ \ \ \ \ \ \ \ \ \ \ \ \
\ \ \ \ \ \ \ \ \ \ \ \ \ \ \ \ \ \ \ \ \ \ \ \ \ \ \ \ \ \ \ \ \ \ \ \ \ \
\ \ \ \ \ \ \ \ \ \ \ \ \ \ \ \ \ \ \ \ \ \ \ \ \ \ \ \ \ \ \ \ \ \ \ \ \ \
\ \ \ \ \ \ \ \ \ \ \ \ \ \ \ \ \ \ \ \ \ \ \ \ \ \ \ \ \ \ \ \ \ \ \ \ \ \
\ \ \ \ \ \ \ \ \ \ \ \ \ \ \ \ \ \ \ \ \ \ \ \ \ \ \ \ \ \ \ \ \ \ \ \ \ \
\ \ \ \ \ \ \ \ \ \ \ \ \ \ \ \ \ \ \ \ \ \ \ \ \ \ \ \ \ \ \ \ \ \ \ \ \ \
\ \ \ \ \ \ \ \ \ \ \ \ \ \ \ \ \ \ \ \ \ \ \ \ \ \ \ \ \ \ \ \ \ \ \ \ \ \
\ \ \ \ \ \ \ \ \ \ \ \ \ \ \ \ \ \ \ \ \ \ \ \ \ \ \ \ \ \ \ \ \ \ \ \ \ \
\ \ \ \ \ \ \ \ \ \ \ \ \ \ \ \ \ \ \ \ \ \ \ \ \ \ \ \ \ \ \ \ \ \ \ \ \ \
\ \ \ \ \ \ \ \ \ \ \ \ \ \ \ \ \ \ \ \ \ \ \ \ \ \ \ \ \ \ \ \ \ \ \ \ \ \
\ \ \ \ \ \ \ \ \ \ \ \ \ \ \ \ \ \ \ \ \ \ \ \ \ \ \ \ \ \ \ \ \ \ \ \ \ \
\ \ \ \ \ \ \ \ \ \ \ \ \ \ \ \ \ \ \ \ \ \ \ \ \ \ \ \ \ \ \ \ \ \ \ \ \ \
\ \ \ \ \ \ \ \ \ \ \ \ \ \ \ \ \ \ \ \ \ \ \ \ \ \ \ \ \ \ \ \ \ \ \ \ \ \
\ \ \ \ \ \ \ \ \ \ \ \ \ \ \ \ \ \ \ \ \ \ \ \ \ \ \ \ \ \ \ \ \ \ \ \ \ \
\ \ \ \ \ \ \ \ \ \ \ \ \ \ \ \ \ \ \ \ \ \ \ \ \ \ \ \ \ \ \ \ \ \ \ \ \ \
\ \ \ \ \ \ \ \ \ \ \ \ \ \ \ \ \ \ \ \ \ \ \ \ \ \ \ \ \ \ \ \ \ \ \ \ \ \
\ \ \ \ \ \ \ \ \ \ \ \ \ \ \ \ \ \ \ \ \ \ \ \ \ \ \ \ \ \ \ \ \ \ \ \ \ \
\ \ \ \ \ \ \ \ \ \ \ \ \ \ \ \ \ \ \ \ \ \ \ \ \ \ \ \ \ \ \ \ \ \ \ \ \ \
\ \ \ \ \ \ \ \ \ \ \ \ \ \ \ \ \ \ \ \ \ \ \ \ \ \ \ \ \ \ \ \ \ \ \ \ \ \
\ \ \ \ \ \ \ \ \ \ \ \ \ \ \ \ \ \ \ \ \ \ \ \ \ \ \ \ \ \ \ \ \ \ \ \ \ \
\ \ \ \ \ \ \ \ \ \ \ \ \ \ \ \ \ \ \ \ \ \ \ \ \ \ \ \ \ \ \ \ \ \ \ \ \ \
\ \ \ \ \ \ \ \ \ \ \ \ \ \ \ \ \ \ \ \ \ \ \ \ \ \ \ \ \ \ \ \ \ \ \ \ \ \
\ \ \ \ \ \ \ \ \ \ \ \ \ \ \ \ \ \ \ \ \ \ \ \ \ \ \ \ \ \ \ \ \ \ \ \ \ \
\ \ \ \ \ \ \ \ \ \ \ \ \ \ \ \ \ \ \ \ \ \ \ \ \ \ \ \ \ \ \ \ \ \ \ \ \ \
\ \ \ \ \ \ \ \ \ \ \ \ \ \ \ \ \ \ \ \ \ \ \ \ \ \ \ \ \ \ \ \ \ \ \ \ \ \
\ \ \ \ \ \ \ \ \ \ \ \ \ \ \ \ \ \ \ \ \ \ \ \ \ \ \ \ \ \ \ \ \ \ \ \ \ \
\ \ \ \ \ \ \ \ \ \ \ \ \ \ \ \ \ \ \ \ \ \ \ \ \ \ \ \ \ \ \ \ \ \ \ \ \ \
\ \ \ \ \ \ \ \ \ \ \ \ \ \ \ \ \ \ \ \ \ \ \ \ \ \ \ \ \ \ \ \ \ \ \ \ \ \
\ \ \ \ \ \ \ \ \ \ \ \ \ \ \ \ \ \ \ \ \ \ \ \ \ \ \ \ \ \ \ \ \ \ \ \ \ \
\ \ \ \ \ \ \ \ \ \ \ \ \ \ \ \ \ \ \ \ \ \ \ \ \ \ \ \ \ \ \ \ \ \ \ \ \ \
\ \ \ \ \ \ \ \ \ \ \ \ \ \ \ \ \ \ \ \ \ \ \ \ \ \ \ \ \ \ \ \ \ \ \ \ \ \
\ \ \ \ \ \ \ \ \ \ \ \ \ \ \ \ \ \ \ \ \ \ \ \ \ \ \ \ \ \ \ \ \ \ \ \ \ \
\ \ \ \ \ \ \ \ \ \ \ \ \ \ \ \ \ \ \ \ \ \ \ \ \ \ \ \ \ \ \ \ \ \ \ \ \ \
\ \ \ \ \ \ \ \ \ \ \ \ \ \ \ \ \ \ \ \ \ \ \ \ \ \ \ \ \ \ \ \ \ \ \ \ \ \
\ \ \ \ \ \ \ \ \ \ \ \ \ \ \ \ \ \ \ \ \ \ \ \ \ \ \ \ \ \ \ \ \ \ \ \ \ \
\ \ \ \ \ \ \ \ \ \ \ \ \ \ \ \ \ \ \ \ \ \ \ \ \ \ \ \ \ \ \ \ \ \ \ \ \ \
\ \ \ \ \ \ \ \ \ \ \ \ \ \ \ \ \ \ \ \ \ \ \ \ \ \ \ \ \ \ \ \ \ \ \ \ \ \
\ \ \ \ \ \ \ \ \ \ \ \ \ \ \ \ \ \ \ \ \ \ \ \ \ \ \ \ \ \ \ \ \ \ \ \ \ \
\ \ \ \ \ \ \ \ \ \ \ \ \ \ \ \ \ \ \ \ \ \ \ \ \ \ \ \ \ \ \ \ \ \ \ \ \ \
\ \ \ \ \ \ \ \ \ \ \ \ \ \ \ \ \ \ \ \ \ \ \ \ \ \ \ \ \ \ \ \ \ \ \ \ \ \
\ \ \ \ \ \ \ \ \ \ \ \ \ \ \ \ \ \ \ \ \ \ \ \ \ \ \ \ \ \ \ \ \ \ \ \ \ \
\ \ \ \ \ \ \ \ \ \ \ \ \ \ \ \ \ \ \ \ \ \ \ \ \ \ \ \ \ \ \ \ \ \ \ \ \ \
\ \ \ \ \ \ \ \ \ \ \ \ \ \ \ \ \ \ \ \ \ \ \ \ \ \ \ \ \ \ \ \ \ \ \ \ \ \
\ \ \ \ \ \ \ \ \ \ \ \ \ \ \ \ \ \ \ \ \ \ \ \ \ \ \ \ \ \ \ \ \ \ \ \ \ \
\ \ \ \ \ \ \ \ \ \ \ \ \ \ \ \ \ \ \ \ \ \ \ \ \ \ \ \ \ \ \ \ \ \ \ \ \ \
\ \ \ \ \ \ \ \ \ \ \ \ \ \ \ \ \ \ \ \ \ \ \ \ \ \ \ \ \ \ \ \ \ \ \ \ \ \
\ \ \ \ \ \ \ \ \ \ \ \ \ \ \ \ \ \ \ \ \ \ \ \ \ \ \ \ \ \ \ \ \ \ \ \ \ \
\ \ \ \ \ \ \ \ \ \ \ \ \ \ \ \ \ \ \ \ \ \ \ \ \ \ \ \ \ \ \ \ \ \ \ \ \ \
\ \ \ \ \ \ \ \ \ \ \ \ \ \ \ \ \ \ \ \ \ \ \ \ \ \ \ \ \ \ \ \ \ \ \ \ \ \
\ \ \ \ \ \ \ \ \ \ \ \ \ \ \ \ \ \ \ \ \ \ \ \ \ \ \ \ \ \ \ \ \ \ \ \ \ \
\ \ \ \ \ \ \ \ \ \ \ \ \ \ \ \ \ \ \ \ \ \ \ \ \ \ \ \ \ \ \ \ \ \ \ \ \ \
\ \ \ \ \ \ \ \ \ \ \ \ \ \ \ \ \ \ \ \ \ \ \ \ \ \ \ \ \ \ \ \ \ \ \ \ \ \
\ \ \ \ \ \ \ \ \ \ \ \ \ \ \ \ \ \ \ \ \ \ \ \ \ \ \ \ \ \ \ \ \ \ \ \ \ \
\ \ \ \ \ \ \ \ \ \ \ \ \ \ \ \ \ \ \ \ \ \ \ \ \ \ \ \ \ \ \ \ \ \ \ \ \ \
\ \ \ \ \ \ \ \ \ \ \ \ \ \ \ \ \ \ \ \ \ \ \ \ \ \ \ \ \ \ \ \ \ \ \ \ \ \
\ \ \ \ \ \ \ \ \ \ \ \ \ \ \ \ \ \ \ \ \ \ \ \ \ \ \ \ \ \ \ \ \ \ \ \ \ \
\ \ \ \ \ \ \ \ \ \ \ \ \ \ \ \ \ \ \ \ \ \ \ \ \ \ \ \ \ \ \ \ \ \ \ \ \ \
\ \ \ \ \ \ \ \ \ \ \ \ \ \ \ \ \ \ \ \ \ \ \ \ \ \ \ \ \ \ \ \ \ \ \ \ \ \
\ \ \ \ \ \ \ \ \ \ \ \ \ \ \ \ \ \ \ \ \ \ \ \ \ \ \ \ \ \ \ \ \ \ \ \ \ \
\ \ \ \ \ \ \ \ \ \ \ \ \ \ \ \ \ \ \ \ \ \ \ \ \ \ \ \ \ \ \ \ \ \ \ \ \ \
\ \ \ \ \ \ \ \ \ \ \ \ \ \ \ \ \ \ \ \ \ \ \ \ \ \ \ \ \ \ \ \ \ \ \ \ \ \
\ \ \ \ \ \ \ \ \ \ \ \ \ \ \ \ \ \ \ \ \ \ \ \ \ \ \ \ \ \ \ \ \ \ \ \ \ \
\ \ \ \ \ \ \ \ \ \ \ \ \ \ \ \ \ \ \ \ \ \ \ \ \ \ \ \ \ \ \ \ \ \ \ \ \ \
\ \ \ \ \ \ \ \ \ \ \ \ \ \ \ \ \ \ \ \ \ \ \ \ \ \ \ \ \ \ \ \ \ \ \ \ \ \
\ \ \ \ \ \ \ \ \ \ \ \ \ \ \ \ \ \ \ \ \ \ \ \ \ \ \ \ \ \ \ \ \ \ \ \ \ \
\ \ \ \ \ \ \ \ \ \ \ \ \ \ \ \ \ \ \ \ \ \ \ \ \ \ \ \ \ \ \ \ \ \ \ \ \ \
\ \ \ \ \ \ \ \ \ \ \ \ \ \ \ \ \ \ \ \ \ \ \ \ \ \ \ \ \ \ \ \ \ \ \ \ \ \
\ \ \ \ \ \ \ \ \ \ \ \ \ \ \ \ \ \ \ \ \ \ \ \ \ \ \ \ \ \ \ \ \ \ \ \ \ \
\ \ \ \ \ \ \ \ \ \ \ \ \ \ \ \ \ \ \ \ \ \ \ \ \ \ \ \ \ \ \ \ \ \ \ \ \ \
\ \ \ \ \ \ \ \ \ \ \ \ \ \ \ \ \ \ \ \ \ \ \ \ \ \ \ \ \ \ \ \ \ \ \ \ \ \
\ \ \ \ \ \ \ \ \ \ \ \ \ \ \ \ \ \ \ \ \ \ \ \ \ \ \ \ \ \ \ \ \ \ \ \ \ \
\ \ \ \ \ \ \ \ \ \ \ \ \ \ \ \ \ \ \ \ \ \ \ \ \ \ \ \ \ \ \ \ \ \ \ \ \ \
\ \ \ \ \ \ \ \ \ \ \ \ \ \ \ \ \ \ \ \ \ \ \ \ \ \ \ \ \ \ \ \ \ \ \ \ \ \
\ \ \ \ \ \ \ \ \ \ \ \ \ \ \ \ \ \ \ \ \ \ \ \ \ \ \ \ \ \ \ \ \ \ \ \ \ \
\ \ \ \ \ \ \ \ \ \ \ \ \ \ \ \ \ \ \ \ \ \ \ \ \ \ \ \ \ \ \ \ \ \ \ \ \ \
\ \ \ \ \ \ \ \ \ \ \ \ \ \ \ \ \ \ \ \ \ \ \ \ \ \ \ \ \ \ \ \ \ \ \ \ \ \
\ \ \ \ \ \ \ \ \ \ \ \ \ \ \ \ \ \ \ \ \ \ \ \ \ \ \ \ \ \ \ \ \ \ \ \ \ \
\ \ \ \ \ \ \ \ \ \ \ \ \ \ \ \ \ \ \ \ \ \ \ \ \ \ \ \ \ \ \ \ \ \ \ \ \ \
\ \ \ \ \ \ \ \ \ \ \ \ \ \ \ \ \ \ \ \ \ \ \ \ \ \ \ \ \ \ \ \ \ \ \ \ \ \
\ \ \ \ \ \ \ \ \ \ \ \ \ \ \ \ \ \ \ \ \ \ \ \ \ \ \ \ \ \ \ \ \ \ \ \ \ \
\ \ \ \ \ \ \ \ \ \ \ \ \ \ \ \ \ \ \ \ \ \ \ \ \ \ \ \ \ \ \ \ \ \ \ \ \ \
\ \ \ \ \ \ \ \ \ \ \ \ \ \ \ \ \ \ \ \ \ \ \ \ \ \ \ \ \ \ \ \ \ \ \ \ \ \
\ \ \ \ \ \ \ \ \ \ \ \ \ \ \ \ \ \ \ \ \ \ \ \ \ \ \ \ \ \ \ \ \ \ \ \ \ \
\ \ \ \ \ \ \ \ \ \ \ \ \ \ \ \ \ \ \ \ \ \ \ \ \ \ \ \ \ \ \ \ \ \ \ \ \ \
\ \ \ \ \ \ \ \ \ \ \ \ \ \ \ \ \ \ \ \ \ \ \ \ \ \ \ \ \ \ \ \ \ \ \ \ \ \
\ \ \ \ \ \ \ \ \ \ \ \ \ \ \ \ \ \ \ \ \ \ \ \ \ \ \ \ \ \ \ \ \ \ \ \ \ \
\ \ \ \ \ \ \ \ \ \ \ \ \ \ \ \ \ \ \ \ \ \ \ \ \ \ \ \ \ \ \ \ \ \ \ \ \ \
\ \ \ \ \ \ \ \ \ \ \ \ \ \ \ \ \ \ \ \ \ \ \ \ \ \ \ \ \ \ \ \ \ \ \ \ \ \
\ \ \ \ \ \ \ \ \ \ \ \ \ \ \ \ \ \ \ \ \ \ \ \ \ \ \ \ \ \ \ \ \ \ \ \ \ \
\ \ \ \ \ \ \ \ \ \ \ \ \ \ \ \ \ \ \ \ \ \ \ \ \ \ \ \ \ \ \ \ \ \ \ \ \ \
\ \ \ \ \ \ \ \ \ \ \ \ \ \ \ \ \ \ \ \ \ \ \ \ \ \ \ \ \ \ \ \ \ \ \ \ \ \
\ \ \ \ \ \ \ \ \ \ \ \ \ \ \ \ \ \ \ \ \ \ \ \ \ \ \ \ \ \ \ \ \ \ \ \ \ \
\ \ \ \ \ \ \ \ \ \ \ \ \ \ \ \ \ \ \ \ \ \ \ \ \ \ \ \ \ \ \ \ \ \ \ \ \ \
\ \ \ \ \ \ \ \ \ \ \ \ \ \ \ \ \ \ \ \ \ \ \ \ \ \ \ \ \ \ \ \ \ \ \ \ \ \
\ \ \ \ \ \ \ \ \ \ \ \ \ \ \ \ \ \ \ \ \ \ \ \ \ \ \ \ \ \ \ \ \ \ \ \ \ \
\ \ \ \ \ \ \ \ \ \ \ \ \ \ \ \ \ \ \ \ \ \ \ \ \ \ \ \ \ \ \ \ \ \ \ \ \ \
\ \ \ \ \ \ \ \ \ \ \ \ \ \ \ \ \ \ \ \ \ \ \ \ \ \ \ \ \ \ \ \ \ \ \ \ \ \
\ \ \ \ \ \ \ \ \ \ \ \ \ \ \ \ \ \ \ \ \ \ \ \ \ \ \ \ \ \ \ \ \ \ \ \ \ \
\ \ \ \ \ \ \ \ \ \ \ \ \ \ \ \ \ \ \ \ \ \ \ \ \ \ \ \ \ \ \ \ \ \ \ \ \ \
\ \ \ \ \ \ \ \ \ \ \ \ \ \ \ \ \ \ \ \ \ \ \ \ \ \ \ \ \ \ \ \ \ \ \ \ \ \
\ \ \ \ \ \ \ \ \ \ \ \ \ \ \ \ \ \ \ \ \ \ \ \ \ \ \ \ \ \ \ \ \ \ \ \ \ \
\ \ \ \ \ \ \ \ \ \ \ \ \ \ \ \ \ \ \ \ \ \ \ \ \ \ \ \ \ \ \ \ \ \ \ \ \ \
\ \ \ \ \ \ \ \ \ \ \ \ \ \ \ \ \ \ \ \ \ \ \ \ \ \ \ \ \ \ \ \ \ \ \ \ \ \
\ \ \ \ \ \ \ \ \ \ \ \ \ \ \ \ \ \ \ \ \ \ \ \ \ \ \ \ \ \ \ \ \ \ \ \ \ \
\ \ \ \ \ \ \ \ \ \ \ \ \ \ \ \ \ \ \ \ \ \ \ \ \ \ \ \ \ \ \ \ \ \ \ \ \ \
\ \ \ \ \ \ \ \ \ \ \ \ \ \ \ \ \ \ \ \ \ \ \ \ \ \ \ \ \ \ \ \ \ \ \ \ \ \
\ \ \ \ \ \ \ \ \ \ \ \ \ \ \ \ \ \ \ \ \ \ \ \ \ \ \ \ \ \ \ \ \ \ \ \ \ \
\ \ \ \ \ \ \ \ \ \ \ \ \ \ \ \ \ \ \ \ \ \ \ \ \ \ \ \ \ \ \ \ \ \ \ \ \ \
\ \ \ \ \ \ \ \ \ \ \ \ \ \ \ \ \ \ \ \ \ \ \ \ \ \ \ \ \ \ \ \ \ \ \ \ \ \
\ \ \ \ \ \ \ \ \ \ \ \ \ \ \ \ \ \ \ \ \ \ \ \ \ \ \ \ \ \ \ \ \ \ \ \ \ \
\ \ \ \ \ \ \ \ \ \ \ \ \ \ \ \ \ \ \ \ \ \ \ \ \ \ \ \ \ \ \ \ \ \ \ \ \ \
\ \ \ \ \ \ \ \ \ \ \ \ \ \ \ \ \ \ \ \ \ \ \ \ \ \ \ \ \ \ \ \ \ \ \ \ \ \
\ \ \ \ \ \ \ \ \ \ \ \ \ \ \ \ \ \ \ \ \ \ \ \ \ \ \ \ \ \ \ \ \ \ \ \ \ \
\ \ \ \ \ \ \ \ \ \ \ \ \ \ \ \ \ \ \ \ \ \ \ \ \ \ \ \ \ \ \ \ \ \ \ \ \ \
\ \ \ \ \ \ \ \ \ \ \ \ \ \ \ \ \ \ \ \ \ \ \ \ \ \ \ \ \ \ \ \ \ \ \ \ \ \
\ \ \ \ \ \ \ \ \ \ \ \ \ \ \ \ \ \ \ \ \ \ \ \ \ \ \ \ \ \ \ \ \ \ \ \ \ \
\ \ \ \ \ \ \ \ \ \ \ \ \ \ \ \ \ \ \ \ \ \ \ \ \ \ \ \ \ \ \ \ \ \ \ \ \ \
\ \ \ \ \ \ \ \ \ \ \ \ \ \ \ \ \ \ \ \ \ \ \ \ \ \ \ \ \ \ \ \ \ \ \ \ \ \
\ \ \ \ \ \ \ \ \ \ \ \ \ \ \ \ \ \ \ \ \ \ \ \ \ \ \ \ \ \ \ \ \ \ \ \ \ \
\ \ \ \ \ \ \ \ \ \ \ \ \ \ \ \ \ \ \ \ \ \ \ \ \ \ \ \ \ \ \ \ \ \ \ \ \ \
\ \ \ \ \ \ \ \ \ \ \ \ \ \ \ \ \ \ \ \ \ \ \ \ \ \ \ \ \ \ \ \ \ \ \ \ \ \
\ \ \ \ \ \ \ \ \ \ \ \ \ \ \ \ \ \ \ \ \ \ \ \ \ \ \ \ \ \ \ \ \ \ \ \ \ \
\ \ \ \ \ \ \ \ \ \ \ \ \ \ \ \ \ \ \ \ \ \ \ \ \ \ \ \ \ \ \ \ \ \ \ \ \ \
\ \ \ \ \ \ \ \ \ \ \ \ \ \ \ \ \ \ \ \ \ \ \ \ \ \ \ \ \ \ \ \ \ \ \ \ \ \
\ \ \ \ \ \ \ \ \ \ \ \ \ \ \ \ \ \ \ \ \ \ \ \ \ \ \ \ \ \ \ \ \ \ \ \ \ \
\ \ \ \ \ \ \ \ \ \ \ \ \ \ \ \ \ \ \ \ \ \ \ \ \ \ \ \ \ \ \ \ \ \ \ \ \ \
\ \ \ \ \ \ \ \ \ \ \ \ \ \ \ \ \ \ \ \ \ \ \ \ \ \ \ \ \ \ \ \ \ \ \ \ \ \
\ \ \ \ \ \ \ \ \ \ \ \ \ \ \ \ \ \ \ \ \ \ \ \ \ \ \ \ \ \ \ \ \ \ \ \ \ \
\ \ \ \ \ \ \ \ \ \ \ \ \ \ \ \ \ \ \ \ \ \ \ \ \ \ \ \ \ \ \ \ \ \ \ \ \ \
\ \ \ \ \ \ \ \ \ \ \ \ \ \ \ \ \ \ \ \ \ \ \ \ \ \ \ \ \ \ \ \ \ \ \ \ \ \
\ \ \ \ \ \ \ \ \ \ \ \ \ \ \ \ \ \ \ \ \ \ \ \ \ \ \ \ \ \ \ \ \ \ \ \ \ \
\ \ \ \ \ \ \ \ \ \ \ \ \ \ \ \ \ \ \ \ \ \ \ \ \ \ \ \ \ \ \ \ \ \ \ \ \ \
\ \ \ \ \ \ \ \ \ \ \ \ \ \ \ \ \ \ \ \ \ \ \ \ \ \ \ \ \ \ \ \ \ \ \ \ \ \
\ \ \ \ \ \ \ \ \ \ \ \ \ \ \ \ \ \ \ \ \ \ \ \ \ \ \ \ \ \ \ \ \ \ \ \ \ \
\ \ \ \ \ \ \ \ \ \ \ \ \ \ \ \ \ \ \ \ \ \ \ \ \ \ \ \ \ \ \ \ \ \ \ \ \ \
\ \ \ \ \ \ \ \ \ \ \ \ \ \ \ \ \ \ \ \ \ \ \ \ \ \ \ \ \ \ \ \ \ \ \ \ \ \
\ \ \ \ \ \ \ \ \ \ \ \ \ \ \ \ \ \ \ \ \ \ \ \ \ \ \ \ \ \ \ \ \ \ \ \ \ \
\ \ \ \ \ \ \ \ \ \ \ \ \ \ \ \ \ \ \ \ \ \ \ \ \ \ \ \ \ \ \ \ \ \ \ \ \ \
\ \ \ \ \ \ \ \ \ \ \ \ \ \ \ \ \ \ \ \ \ \ \ \ \ \ \ \ \ \ \ \ \ \ \ \ \ \
\ \ \ \ \ \ \ \ \ \ \ \ \ \ \ \ \ \ \ \ \ \ \ \ \ \ \ \ \ \ \ \ \ \ \ \ \ \
\ \ \ \ \ \ \ \ \ \ \ \ \ \ \ \ \ \ \ \ \ \ \ \ \ \ \ \ \ \ \ \ \ \ \ \ \ \
\ \ \ \ \ \ \ \ \ \ \ \ \ \ \ \ \ \ \ \ \ \ \ \ \ \ \ \ \ \ \ \ \ \ \ \ \ \
\ \ \ \ \ \ \ \ \ \ \ \ \ \ \ \ \ \ \ \ \ \ \ \ \ \ \ \ \ \ \ \ \ \ \ \ \ \
\ \ \ \ \ \ \ \ \ \ \ \ \ \ \ \ \ \ \ \ \ \ \ \ \ \ \ \ \ \ \ \ \ \ \ \ \ \
\ \ \ \ \ \ \ \ \ \ \ \ \ \ \ \ \ \ \ \ \ \ \ \ \ \ \ \ \ \ \ \ \ \ \ \ \ \
\ \ \ \ \ \ \ \ \ \ \ \ \ \ \ \ \ \ \ \ \ \ \ \ \ \ \ \ \ \ \ \ \ \ \ \ \ \
\ \ \ \ \ \ \ \ \ \ \ \ \ \ \ \ \ \ \ \ \ \ \ \ \ \ \ \ \ \ \ \ \ \ \ \ \ \
\ \ \ \ \ \ \ \ \ \ \ \ \ \ \ \ \ \ \ \ \ \ \ \ \ \ \ \ \ \ \ \ \ \ \ \ \ \
\ \ \ \ \ \ \ \ \ \ \ \ \ \ \ \ \ \ \ \ \ \ \ \ \ \ \ \ \ \ \ \ \ \ \ \ \ \
\ \ \ \ \ \ \ \ \ \ \ \ \ \ \ \ \ \ \ \ \ \ \ \ \ \ \ \ \ \ \ \ \ \ \ \ \ \
\ \ \ \ \ \ \ \ \ \ \ \ \ \ \ \ \ \ \ \ \ \ \ \ \ \ \ \ \ \ \ \ \ \ \ \ \ \
\ \ \ \ \ \ \ \ \ \ \ \ \ \ \ \ \ \ \ \ \ \ \ \ \ \ \ \ \ \ \ \ \ \ \ \ \ \
\ \ \ \ \ \ \ \ \ \ \ \ \ \ \ \ \ \ \ \ \ \ \ \ \ \ \ \ \ \ \ \ \ \ \ \ \ \
\ \ \ \ \ \ \ \ \ \ \ \ \ \ \ \ \ \ \ \ \ \ \ \ \ \ \ \ \ \ \ \ \ \ \ \ \ \
\ \ \ \ \ \ \ \ \ \ \ \ \ \ \ \ \ \ \ \ \ \ \ \ \ \ \ \ \ \ \ \ \ \ \ \ \ \
\ \ \ \ \ \ \ \ \ \ \ \ \ \ \ \ \ \ \ \ \ \ \ \ \ \ \ \ \ \ \ \ \ \ \ \ \ \
\ \ \ \ \ \ \ \ \ \ \ \ \ \ \ \ \ \ \ \ \ \ \ \ \ \ \ \ \ \ \ \ \ \ \ \ \ \
\ \ \ \ \ \ \ \ \ \ \ \ \ \ \ \ \ \ \ \ \ \ \ \ \ \ \ \ \ \ \ \ \ \ \ \ \ \
\ \ \ \ \ \ \ \ \ \ \ \ \ \ \ \ \ \ \ \ \ \ \ \ \ \ \ \ \ \ \ \ \ \ \ \ \ \
\ \ \ \ \ \ \ \ \ \ \ \ \ \ \ \ \ \ \ \ \ \ \ \ \ \ \ \ \ \ \ \ \ \ \ \ \ \
\ \ \ \ \ \ \ \ \ \ \ \ \ \ \ \ \ \ \ \ \ \ \ \ \ \ \ \ \ \ \ \ \ \ \ \ \ \
\ \ \ \ \ \ \ \ \ \ \ \ \ \ \ \ \ \ \ \ \ \ \ \ \ \ \ \ \ \ \ \ \ \ \ \ \ \
\ \ \ \ \ \ \ \ \ \ \ \ \ \ \ \ \ \ \ \ \ \ \ \ \ \ \ \ \ \ \ \ \ \ \ \ \ \
\ \ \ \ \ \ \ \ \ \ \ \ \ \ \ \ \ \ \ \ \ \ \ \ \ \ \ \ \ \ \ \ \ \ \ \ \ \
\ \ \ \ \ \ \ \ \ \ \ \ \ \ \ \ \ \ \ \ \ \ \ \ \ \ \ \ \ \ \ \ \ \ \ \ \ \
\ \ \ \ \ \ \ \ \ \ \ \ \ \ \ \ \ \ \ \ \ \ \ \ \ \ \ \ \ \ \ \ \ \ \ \ \ \
\ \ \ \ \ \ \ \ \ \ \ \ \ \ \ \ \ \ \ \ \ \ \ \ \ \ \ \ \ \ \ \ \ \ \ \ \ \
\ \ \ \ \ \ \ \ \ \ \ \ \ \ \ \ \ \ \ \ \ \ \ \ \ \ \ \ \ \ \ \ \ \ \ \ \ \
\ \ \ \ \ \ \ \ \ \ \ \ \ \ \ \ \ \ \ \ \ \ \ \ \ \ \ \ \ \ \ \ \ \ \ \ \ \
\ \ \ \ \ \ \ \ \ \ \ \ \ \ \ \ \ \ \ \ \ \ \ \ \ \ \ \ \ \ \ \ \ \ \ \ \ \
\ \ \ \ \ \ \ \ \ \ \ \ \ \ \ \ \ \ \ \ \ \ \ \ \ \ \ \ \ \ \ \ \ \ \ \ \ \
\ \ \ \ \ \ \ \ \ \ \ \ \ \ \ \ \ \ \ \ \ \ \ \ \ \ \ \ \ \ \ \ \ \ \ \ \ \
\ \ \ \ \ \ \ \ \ \ \ \ \ \ \ \ \ \ \ \ \ \ \ \ \ \ \ \ \ \ \ \ \ \ \ \ \ \
\ \ \ \ \ \ \ \ \ \ \ \ \ \ \ \ \ \ \ \ \ \ \ \ \ \ \ \ \ \ \ \ \ \ \ \ \ \
\ \ \ \ \ \ \ \ \ \ \ \ \ \ \ \ \ \ \ \ \ \ \ \ \ \ \ \ \ \ \ \ \ \ \ \ \ \
\ \ \ \ \ \ \ \ \ \ \ \ \ \ \ \ \ \ \ \ \ \ \ \ \ \ \ \ \ \ \ \ \ \ \ \ \ \
\ \ \ \ \ \ \ \ \ \ \ \ \ \ \ \ \ \ \ \ \ \ \ \ \ \ \ \ \ \ \ \ \ \ \ \ \ \
\ \ \ \ \ \ \ \ \ \ \ \ \ \ \ \ \ \ \ \ \ \ \ \ \ \ \ \ \ \ \ \ \ \ \ \ \ \
\ and%
\[
K_{1/3+2\,j-2\,i}(\zeta )=K_{1/3}(\zeta )\sum_{p=0}^{i-j-1}\tbinom{i-j-1+p}{%
2p}\tfrac{\Gamma (1/3-i+j+p)}{\Gamma (1/3-i+j-p)}\left( \tfrac{2}{\zeta }%
\right) ^{2p} 
\]%
\[
-\left( \tfrac{2}{\zeta }\right) K_{2/3}(\zeta )\sum_{p=0}^{i-j-1}\tbinom{%
i-j+p}{2p+1}\tfrac{\Gamma (4/3-i+j+p)}{\Gamma (1/3-i+j-p)}\left( \tfrac{2}{%
\zeta }\right) ^{2p}. 
\]%
Using these results we get for the Bell sum of odd degree 
\[
\sum_{i=0}^{2\,j+1}\tbinom{2\,j+1}{i}K\,_{-1/3-2\,\,j+2\,i}(\zeta
)=K_{1/3}(\zeta ){\Huge [}\tbinom{2j+1}{j}+\sum_{p=0}^{j}\left( \tfrac{2}{%
\zeta }\right) ^{2p}\left\{ \sum_{i=0}^{j-p}\tbinom{2j+1}{i+j+1+p}\tbinom{%
i+2p}{2p}\tfrac{\Gamma (5/3+i+2p)}{\Gamma (5/3+i)}\right\} 
\]%
\[
+\sum_{p=0}^{j-1}\left( \tfrac{2}{\zeta }\right) ^{2p}\left\{
\sum_{i=0}^{j-p-1}\tbinom{2j+1}{i}\tbinom{j-i+p}{2p+1}\tfrac{\Gamma
(4/3+j-i+p)}{\Gamma (1/3+j-i-p)}\left( \tfrac{\left( 2p+1\right) }{%
(j-i+p)(1/3+j-i-p)}+\tfrac{1}{3}\left( \tfrac{2}{\zeta }\right) ^{2}\right)
\right\} {\Huge ]} 
\]%
\[
+\left( \tfrac{2}{\zeta }\right) K_{2/3}(\zeta ){\Huge [}\sum_{p=0}^{j}%
\left( \tfrac{2}{\zeta }\right) ^{2p}\left\{ \sum_{i=0}^{j-p}\tbinom{2j+1}{%
i+j+1+p}\tbinom{i+2p+1}{2p+1}\tfrac{\Gamma (5/3+i+2p)}{\Gamma (2/3+i)}%
\right\} 
\]%
\begin{equation}
+\sum_{p=0}^{j-1}\left( \tfrac{2}{\zeta }\right) ^{2p}\left\{
\sum_{i=0}^{j-p-1}\tbinom{2j+1}{i}\tbinom{j-i+p}{2p+1}\tfrac{\Gamma
(4/3+j-i+p)}{\Gamma (1/3+j-i-p)}\right\} {\Huge ]},\text{ \ \ for \ }j>0.
\label{eq18}
\end{equation}%
In the equation above we have interchanged the order of summation in the
double sums and have replaced the ratios of Gamma functions with negative
arguments with ones with positive arguments using the relation $\Gamma
(-z)=-\pi \,\csc (\pi z)/\Gamma (z+1).$ \ The expression in Eq. (18) can be
simplified by writing out the $p=0$ terms which occur in the double sums
whose ranges are $0\leq p\leq j$ and then re-indexing those sums. \ We get
as a final expression%
\[
\sum_{i=0}^{2\,j+1}\tbinom{2\,j+1}{i}K_{-1/3-2\,\,j+2\,i}(\zeta
)=K_{1/3}(\zeta )\left[ 2^{2j}+\tbinom{2j+1}{j}+\sum_{p=0}^{j-1}C_{3}(j,p;%
\zeta )\left( \tfrac{2}{\zeta }\right) ^{2p}\right] 
\]%
\begin{equation}
+\left( \tfrac{2}{\zeta }\right) K_{2/3}(\zeta )\left[ (\tfrac{1}{3}+\tfrac{j%
}{2})2^{2j}+\tfrac{1}{3}(j+1)\tbinom{2j+1}{j}+\sum_{p=0}^{j-1}C_{4}(j,p;%
\zeta )\left( \tfrac{2}{\zeta }\right) ^{2p}\right] ,  \label{eq19}
\end{equation}%
where 
\[
C_{3}(j,p;\zeta )=\sum_{i=0}^{j-1-p}{\LARGE \{}\tbinom{2j+1}{i}\tbinom{j+p-i%
}{2p+1}\tfrac{\Gamma (4/3+j-i+p)}{\Gamma (1/3+j-i-p)}\left[ \tfrac{\left(
1+2p\right) }{(j+p-i)(1/3+j-i-p)}+\tfrac{1}{3}\left( \tfrac{2}{\zeta }%
\right) ^{2}\right] 
\]%
\ \ \ \ \ \ \ \ \ \ \ \ \ \ \ \ \ \ \ \ \ \ \ \ \ \ \ \ \ \ \ \ \ \ \ \ \ \
\ \ \ \ \ \ \ \ \ \ \ \ \ \ \ \ \ \ \ \ \ \ \ \ \ \ \ \ \ \ \ \ \ \ \ \ \ \
\ \ \ \ \ \ \ \ \ \ \ \ \ \ \ \ \ \ \ \ \ \ \ \ \ \ \ \ \ \ \ \ \ \ \ \ \ \
\ \ \ \ \ \ \ \ \ \ \ \ \ \ \ \ \ \ \ \ \ \ \ \ \ \ \ \ \ \ \ \ \ \ \ \ \ \
\ \ \ \ \ \ \ \ \ \ \ \ \ \ \ \ \ \ \ \ \ \ \ \ \ \ \ \ \ \ \ \ \ \ \ \ \ \
\ \ \ \ \ \ \ \ \ \ \ \ \ \ 
\[
+\tbinom{2j+1}{i+j+p+2}\tbinom{i+2p+2}{2p+2}\tfrac{\Gamma (11/3+2p+i)}{%
\Gamma (5/3+i)}\left( \tfrac{2}{\zeta }\right) ^{2}{\Large \}},\text{\ for \ 
}j>0, 
\]%
and%
\[
C_{4}(j,p;\zeta )=\sum_{i=0}^{j-1-p}{\Large \{}\tbinom{2j+1}{i}\tbinom{j+p-i%
}{2p+1}\tfrac{\Gamma (4/3+j-i+p)}{\Gamma (1/3+j-i-p)}+\tbinom{2j+1}{i+j+p+2}%
\tbinom{i+2p+3}{2p+3}\tfrac{\Gamma (11/3+2p+i)}{\Gamma (2/3+i)}\left( \tfrac{%
2}{\zeta }\right) ^{2}{\LARGE \}},\text{\ for \ }j>0. 
\]%
As a result, the Bell term with odd degree in Eq. (12) can be written as%
\[
\sum_{j=1}^{m-1}\left( \tfrac{3\zeta }{2}\right) ^{2\,j+1}\Delta
B_{2m,\,2\,j+1}\sum_{i=0}^{2\,j+1}\tbinom{2\ j+1}{i}K_{-1/3+2\,i-2\,j}(\zeta
)= 
\]%
\[
\left( \tfrac{\zeta }{2}\right) K_{1/3}(\zeta ){\Huge [}\sum_{j=1}^{m-1}%
\left\{ 2^{2j}+\tbinom{2j+1}{j}\right\} 3^{2j+1}\Delta B_{2m,2j+1}\left( 
\tfrac{\zeta }{2}\right) ^{2j} 
\]%
\[
+\sum_{k=1}^{m-1}\left( \sum_{j=k}^{m-1}3^{2j+1}\Delta
B_{2m,2j+1}C_{3}(j,j-k;\zeta )\right) \left( \tfrac{\zeta }{2}\right) ^{2k}%
{\Huge ]} 
\]%
\[
+K_{2/3}(\zeta ){\Huge [}\sum_{j=1}^{m-1}\left\{ (1+\tfrac{3}{2}%
j)\,2^{2j}+(j+1)\tbinom{2j+1}{j}\right\} 3^{2j}\Delta B_{2m,2j+1}\left( 
\tfrac{\zeta }{2}\right) ^{2j} 
\]%
\begin{equation}
+\sum_{k=1}^{m-1}\left( \sum_{j=k}^{m-1}3^{2j+1}\Delta
B_{2m,2j+1}C_{4}(j,j-k;\zeta )\right) \left( \tfrac{\zeta }{2}\right) ^{2k}%
{\Huge ]}.  \label{eq20}
\end{equation}

Where we note that in the case where $m=1$ the sums in Eq. (20) are empty. \
Gathering the results in Eqs. (12, 17, 20) we have for%
\[
\frac{d^{\,\,2m}Ai^{^{\prime }}(z)}{d\,z^{2m}}=-\;\frac{1}{\pi \sqrt{3}%
\,2^{2m}\,\left( \frac{3}{2}\zeta \right) ^{\frac{2}{3}(2m-1)}}\sigma
(m;\zeta ), 
\]%
where%
\[
\sigma (m;\zeta )=K_{1/3}(\zeta ){\Huge [}30\,(4m-7)!!(\tfrac{\zeta }{2}%
)-3^{2m}\sum_{j=0}^{m-1}C_{1}^{(0)}(m,m-1-j)(\tfrac{\zeta }{2})^{2j+1} 
\]%
\[
-\,H(m-2)\sum_{j=1}^{m-1}\left[ 2^{2j}+\tbinom{2j+1}{j}\right] \Delta
B_{2m,2j+1}(\tfrac{3\zeta }{2})^{2j+1} 
\]%
\[
-\,H(m-2)\sum_{q=1}^{m-1}\,(\tfrac{\zeta }{2})^{2q+1}\sum_{j=q}^{m-1}{\huge %
\{}3^{2j}\Delta B_{2m,2j}\,C_{1}^{(0)}(j,j-q)(\tfrac{\zeta }{2}%
)^{2q-1}+3^{2j+1}\Delta B_{2m,2j+1}C_{3}(j,j-q;\zeta ){\huge \}}{\Huge ]} 
\]%
\[
+K_{2/3}(\zeta ){\Huge [}10(4m-7)!!+\tbinom{2m}{m}(\tfrac{3\zeta }{2}%
)^{2m}+3^{2m}\sum_{j=1}^{m}C_{2}(m,m-j;\zeta )(\tfrac{\zeta }{2})^{2j} 
\]%
\[
+\,H(m-2)\sum_{j=1}^{m-1}\left[ \tbinom{2j}{j}\Delta B_{2m,2j}-\left\{ (1+%
\tfrac{3}{2}j)2^{2j}+(j+1)\tbinom{2j+1}{j}\right\} \Delta B_{2m,2j+1}\right]
(\tfrac{3\zeta }{2})^{2j} 
\]%
\begin{equation}
+\,H(m-2)\sum_{q=1}^{m-1}\,(\tfrac{\zeta }{2})^{2q}\sum_{j=q}^{m-1}3^{2j}%
{\LARGE [}\Delta B_{2m,2j}\,C_{2}(j,j-q;\zeta )-3\,\Delta
B_{2m,2j+1}C_{4}(j,j-q;\zeta ){\LARGE ]}{\Huge ]}.  \label{eq21}
\end{equation}%
In the equation below we have combined terms appearing in Eq. (21) noting
that $\Delta B_{2m,1}=-5(4m-7)!!$ and having rewritten the sums in
preparation for the extraction of the sought-after polynomials from that
expression. \ We have 
\[
\sigma (m;\zeta )=-\left( \tfrac{\zeta }{2}\right) K_{1/3}(\zeta ){\Huge [}%
\sum_{j=0}^{m-1}{\huge \{}3^{2m}\,C_{1}^{(0)}(m,m-1-j)+{\Large [}2^{2j}+%
\tbinom{2j+1}{j}{\Large ]}3^{2j+1}\Delta B_{2m,2j+1}{\huge \}}(\tfrac{\zeta 
}{2})^{2j} 
\]%
\[
+H(m-2)\sum_{q=1}^{m-1}\,(\tfrac{\zeta }{2})^{2q-2}\sum_{j=q}^{m-1}3^{2j}%
{\huge \{}C_{1}^{(0)}(j,j-q)\,\Delta B_{2m,2j}\,+3\,\left( {\small \zeta /2}%
\right) ^{2}{\small \cdot }\,C_{3}(j,j-q;\zeta )\,\Delta B_{2m,2j+1}{\huge \}%
}{\Huge ]} 
\]%
\[
+K_{2/3}(\zeta ){\Huge [}10(4m-7)!!+\tbinom{2m}{m}(\tfrac{3\zeta }{2}%
)^{2m}+3^{2m}\sum_{j=1}^{m}C_{2}(m,m-j;\zeta )(\tfrac{\zeta }{2})^{2j} 
\]%
\[
+\,\sum_{j=1}^{m-1}\left[ \tbinom{2j}{j}\Delta B_{2m,2j}-\left\{ (1+\tfrac{3%
}{2}j)2^{2j}+(j+1)\tbinom{2j+1}{j}\right\} \Delta B_{2m,2j+1}\right] (\tfrac{%
3\zeta }{2})^{2j} 
\]%
\begin{equation}
+\,H(m-2)\sum_{q=1}^{m-1}(\tfrac{\zeta }{2})^{2q}\sum_{j=q}^{m-1}3^{2j}%
{\huge \{}C_{2}(j,j-q;\zeta )\,\Delta B_{2m,2j}-3\,C_{4}(j,j-q;\zeta
)\,\Delta B_{2m,2j+1}{\huge \}}{\Huge ]}.  \label{eq22}
\end{equation}%
\ We have%
\[
\mathcal{P}_{2m+1}(z)=\frac{\frac{1}{3}\sigma _{1/3}(m;\zeta )}{2^{2m}\,%
{\small (3\,\zeta \,/\,2)}^{4(m-1)/3}},\text{ \ \ \ \ \ \ \ }Q_{2m+1}(z)=%
\frac{\sigma _{2/3}(m;\zeta )}{2^{2m}\,{\small (3\,\zeta \,/\,2)}^{4m/3}}, 
\]%
where%
\[
\sigma _{1/3}(m;\zeta )=\sum_{j=0}^{m-1}{\huge \{}3^{2m}%
\,C_{1}^{(0)}(m,m-1-j)+{\Large [}2^{2j}+\tbinom{2j+1}{j}{\Large ]}%
3^{2j+1}\Delta B_{2m,2j+1}{\huge \}}(\tfrac{\zeta }{2})^{2j} 
\]%
\begin{equation}
+H(m-2)\sum_{q=1}^{m-1}(\tfrac{\zeta }{2})^{2q-2}\sum_{j=q}^{m-1}3^{2j}%
{\huge \{}C_{1}^{(0)}(j,\,j-q)\,\Delta B_{2m,2j}\,+\,3\,\left( {\small \zeta
\,/2}\right) ^{2}{\small \cdot }\,C_{3}(j,\,j-q;\zeta )\,\Delta B_{2m,2j+1}%
{\huge \}},  \label{eq23}
\end{equation}%
and%
\[
\sigma _{2/3}(m;\zeta )=10(4m-7)!!+\tbinom{2m}{m}(\tfrac{3\zeta }{2}%
)^{2m}+3^{2m}\sum_{j=1}^{m}C_{2}(m,m-j;\zeta )(\tfrac{\zeta }{2})^{2j} 
\]%
\[
+\,H(m-2)\sum_{j=1}^{m-1}\left[ \tbinom{2j}{j}\Delta B_{2m,2j}-\left\{ (1+%
\tfrac{3}{2}j)2^{2j}+(j+1)\tbinom{2j+1}{j}\right\} \Delta B_{2m,2j+1}\right]
(\tfrac{3\zeta }{2})^{2j} 
\]%
\begin{equation}
+\,H(m-2)\sum_{q=1}^{m-1}(\tfrac{\zeta }{2})^{2q}\sum_{j=q}^{m-1}3^{2j}%
{\Huge \{}C_{2}(j,\,j-q;\zeta )\,\Delta B_{2m,2j}-3\,C_{4}(j,\,j-q;\zeta
)\,\Delta B_{2m,2j+1}{\Huge \}}.  \label{eq24}
\end{equation}

In the expressions for $\mathcal{P}_{2m+1}(z)$ and $Q_{2m+1}(z)$ given
above, cancellations of terms with negative powers of $\zeta $ must occur in 
$\sigma _{1/3}/{\small (\zeta \,/\,2)}^{4(m-1)/3}$ and $\sigma _{2/3}/%
{\small (\zeta \,/\,2)}^{4m/3}$ leaving terms with only zero or positive
exponents as is required in the sought-after polynomials. \ To this end, we
rewrite the terms in $\sigma _{1/3}$ and $\sigma _{2/3}$ gathering together $%
\zeta $ terms of the same order. \ In this regard it is helpful to write the
coefficients $C_{2},$ $C_{3},$ $C_{4},$ as 
\[
C_{2}(m,p;\zeta )=C_{2}^{(0)}(m,p)+(2/\zeta )^{2}C_{2}^{(2)}(m,p), 
\]%
\[
C_{3}(m,p;\zeta )=C_{3}^{(0)}(m,p)+(2/\zeta )^{2}C_{3}^{(2)}(m,p), 
\]%
\[
C_{4}(m,p;\zeta )=C_{4}^{(0)}(m,p)+(2/\zeta )^{2}C_{4}^{(2)}(m,p). 
\]

\begin{figure}[!h]
\noindent \begin{centering}
\includegraphics{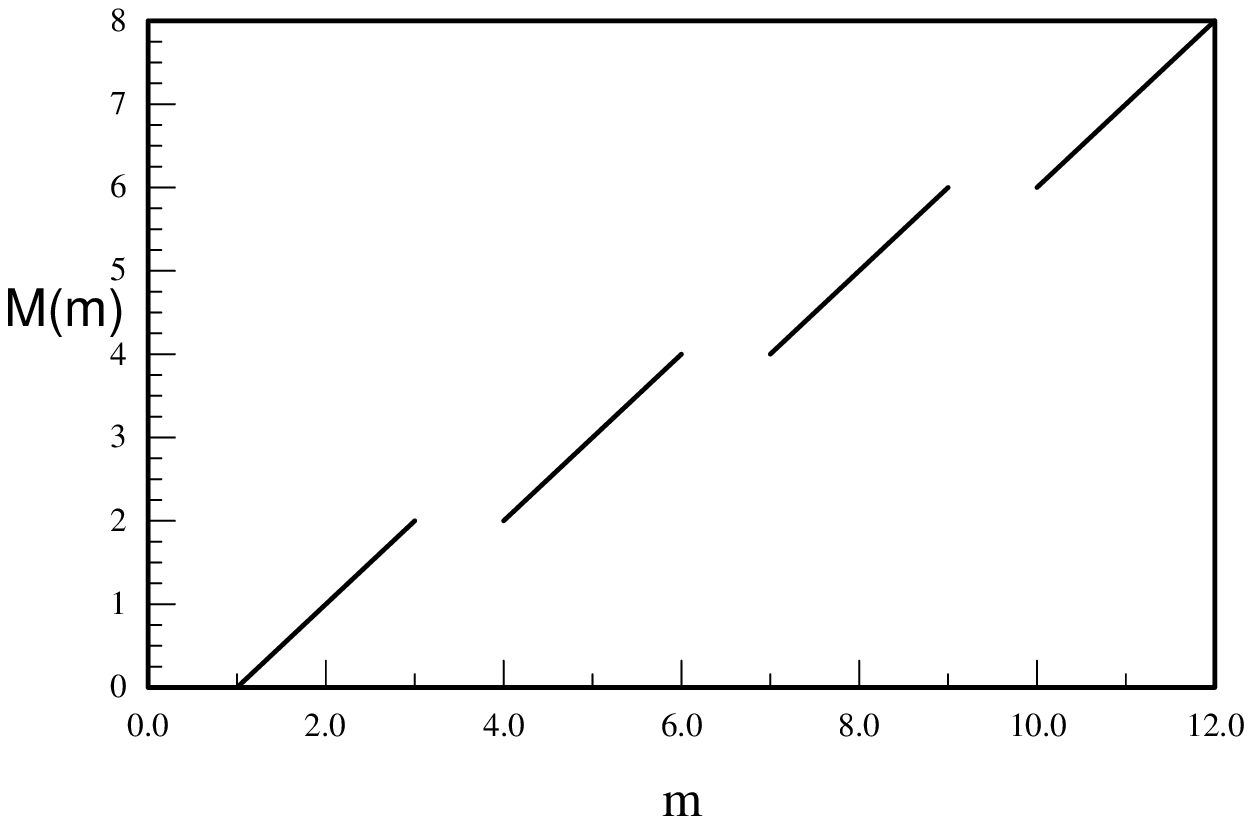}\caption{\label{fig:1}}
\par\end{centering}
\end{figure}

In the case of $m\geq 4,$ we have%
\[
\frac{\sigma _{1/3}(m;\zeta )}{(\zeta /2)^{4(m-1)/3}}=\sum_{q=2}^{m-1}\left[
A(m,q)+D^{(2)}(m,q)\right] (\zeta
/2)^{2[q-2(m-1)/3]}+\sum_{q=2}^{m-2}D^{(0)}(m,q+1)(\zeta
/2)^{2[q-2(m-1)/3]}. 
\]%
where 
\[
A(m,j)=3^{2m}\,C_{1}^{(0)}(m,m-1-j)+{\Large [}2^{2j}+\tbinom{2j+1}{j}{\Large %
]}3^{2j+1}\Delta B_{2m,2j+1}, 
\]%
\[
D^{(0)}(m,q)=\sum_{j=q}^{m-1}3^{2j}{\huge \{}C_{1}^{(0)}(j,\,j-q)\,\Delta
B_{2m,2j}\,+\,3\,C_{3}^{(2)}(j,\,j-q)\,\Delta B_{2m,2j+1}{\huge \}}, 
\]%
\[
D^{(2)}(m,q)=\sum_{j=q}^{m-1}3^{2j+1}\,C_{3}^{(0)}(j,\,j-q)\,\Delta
B_{2m,2j+1}{\huge .} 
\]%
Since the exponent $q-2(m-1)/3\geq 0$ the sums above may be replaced by%
\begin{equation}
\frac{\sigma _{1/3}(m;\zeta )}{(\zeta /2)^{4(m-1)/3}}=\sum_{q=\mathsf{M}%
(m)}^{m-1}\left[ A(m,q)+D^{(2)}(m,q)\right] (\zeta
/2)^{2[q-2(m-1)/3]}+\sum_{q=\mathsf{M}(m)}^{m-2}D^{(0)}(m,q+1)(\zeta
/2)^{2[q-2(m-1)/3]},  \label{eq25}
\end{equation}%
where%
\[
\mathsf{M}(m)=\left\lceil \tfrac{2\,(m-1)}{3}\right\rceil , 
\]%
and $\left\lceil x\right\rceil $ is the ceiling function \cite{Ceil} whose
value is equal to the smallest integer which is greater that or equal to $x.$
A plot of the function $\mathsf{M}(m)$ is shown in Figure \ref{fig:1}.

\begin{figure}[!h]
\noindent \begin{centering}
\includegraphics{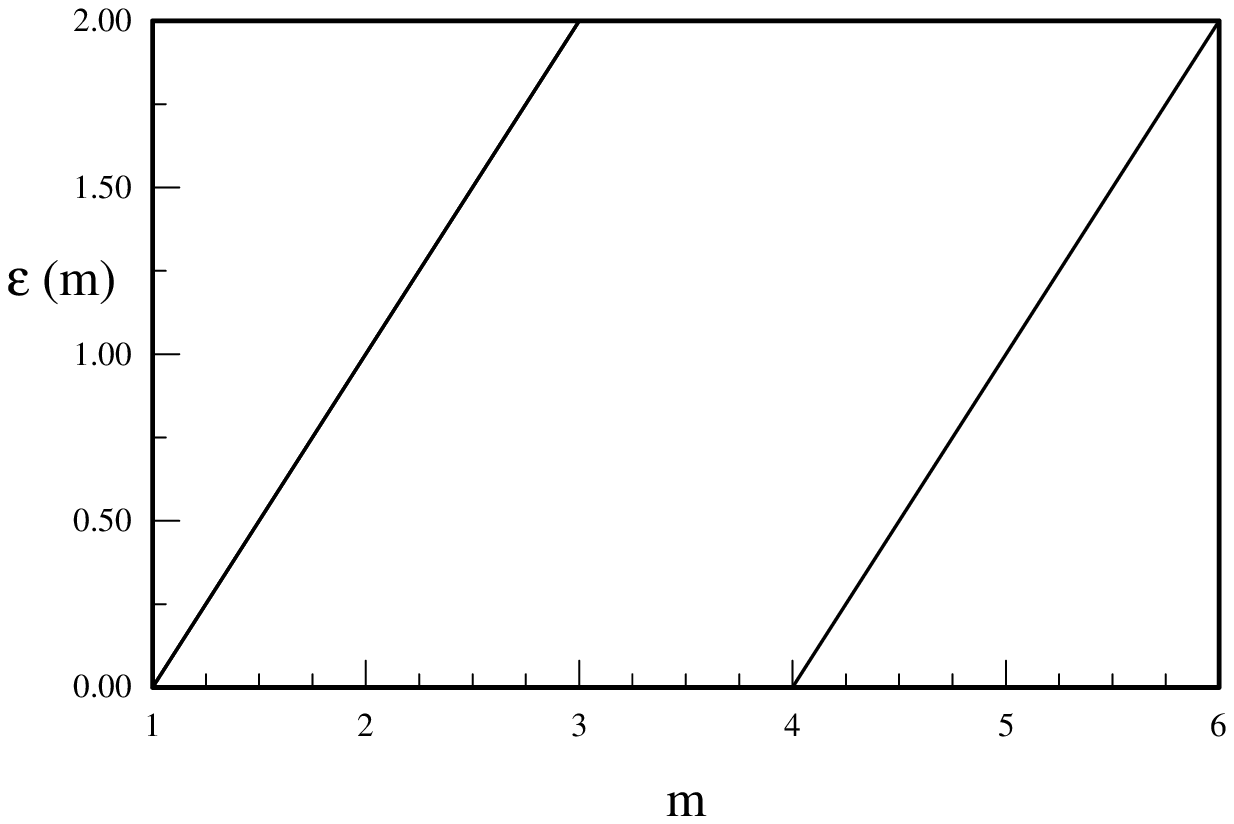}\caption{\label{fig:2}}
\par\end{centering}
\end{figure}

Reindexing the sums in Eq. (25) we have 
\[
\frac{\sigma _{1/3}(m;\zeta )}{(\zeta /2)^{4(m-1)/3}}=\sum_{q=0}^{\mathcal{M(%
}m\mathcal{)}}\left[ A(m,q+\mathsf{M}(m))+D^{(2)}(m,q+\mathsf{M}(m))\right]
(\zeta /2)^{2[q+\mathsf{M}(m)-2(m-1)/3]} 
\]%
\[
+\sum_{q=0}^{\mathcal{M(}m\mathcal{)\,-}1}D^{(0)}(m,q+\mathsf{M}(m)+1)(\zeta
/2)^{2[q+\mathsf{M}(m)-2(m-1)/3]}, 
\]%
where 
\[
\mathcal{M(}m\mathcal{)}=m-1-\mathsf{M}(m). 
\]%
A plot of $\mathcal{M(}m\mathcal{)}$ is seen to be staircase function of $m$
as shown in Figure \ref{fig:2}.

\bigskip

\begin{figure}[!h]
\noindent \begin{centering}
\includegraphics{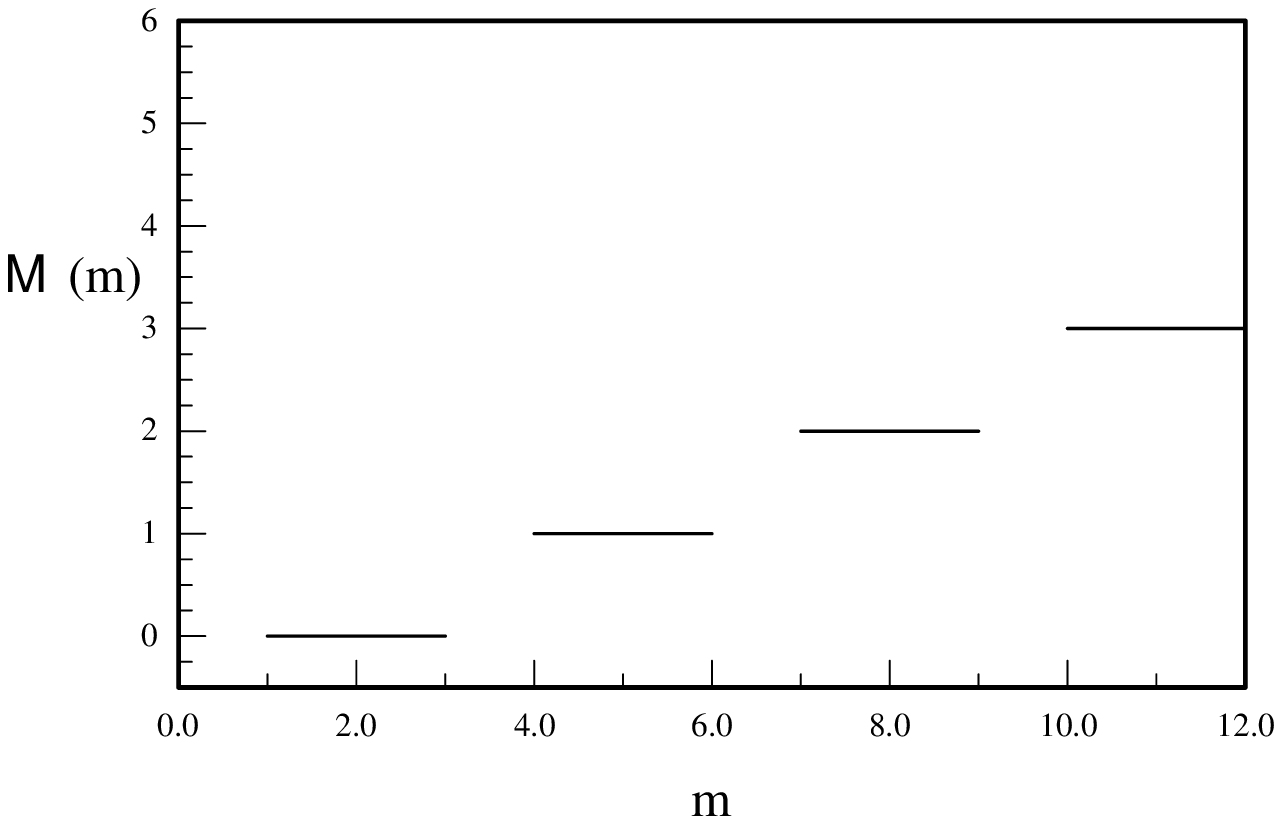}\caption{\label{fig:3}}
\par\end{centering}
\end{figure}

Using the result in Eq. (23) \ we get (with $\zeta =\tfrac{2}{3}z^{\,3/2}$)%
\begin{eqnarray}
\mathcal{P}_{2m+1}(z) &=&\frac{z^{\,\epsilon (m)}}{2^{2m}3^{2\,\mathsf{M}%
(m)+1}}{\Huge \{}\left[ A(m,m-1)+D^{(2)}(m,m-1\,)\right] \left(
z^{3}/9\right) ^{\mathcal{M(}m\mathcal{)}}  \label{eq26} \\
&&\hspace{1.5in}+\sum_{q=0}^{\mathcal{M(}m\mathcal{)\,-}1}\mathbf{C}%
_{1/3}(m,q)\,\left( z^{3}/9\right) ^{q}{\Huge \}},  \nonumber
\end{eqnarray}%
where%
\[
\mathbf{C}_{1/3}(m,q)=A(m,\mathsf{M}(m)+q)+D^{(0)}(m,\mathsf{M}%
(m)+1+q)+D^{(2)}(m,\mathsf{M}(m)+q\,), 
\]%
and where%
\[
\epsilon (m)=3\,\mathsf{M}(m)-2\,(m-1). 
\]%
It should be noted that the exponent $\epsilon (m)$ is a periodic function
of $m$ i.e. $\epsilon (m)=\epsilon (m+4)$ as seen in Figure \ref{fig:3}. \ 

\subsection{General expressions for $\mathcal{P}_{2m+1}(z)$ and $Q_{2m+1}(z)$%
}

The expression in Eq. (26) can be further simplified by evaluating the sums
which are contained in the coefficient of $\left( z^{3}/9\right) ^{\mathcal{%
M(}m\mathcal{)}}$ in conjunction with\ the relation $\Delta
B_{2m,2m-1}=2m^{2}+3m.$ We get%
\[
\mathcal{P}_{2m+1}(z)=m^{2}\,z^{m-1}+z^{\,\epsilon (m)}\,{\small \cdot }%
\sum_{q=0}^{\mathcal{M(}m\mathcal{)\,-}1}\mathbf{C}_{1/3}(m,q)\,\left(
z^{3}/9\right) ^{q}, 
\]%
where the coefficients $\mathbf{C}_{1/3}(m,q)$ are given by 
\[
\mathbf{C}_{1/3}(m,q)=\,\frac{3^{2\mathcal{M(}m\mathcal{)}+1}}{2^{2m}}\,%
\left[ C_{1}^{(0)}(m,\mathcal{M(}m\mathcal{)}-q)+m\,(1+2m/3)%
\,C_{3}^{(0)}(m-1,\mathcal{M(}m\mathcal{)}-q)\right] 
\]%
\begin{equation}
+3^{2q}\,\left[ {\huge \{}2^{2q-2\mathcal{M(}m\mathcal{)}-2}+2^{-2m}\tbinom{2%
\mathsf{M}\mathcal{(}m\mathcal{)}+1+2q}{\mathsf{M}\mathcal{(}m\mathcal{)}+q}%
{\huge \}}\,\Delta B_{2m,\,2\mathsf{M}\mathcal{(}m\mathcal{)}%
+1+2q}\,+2^{-2m}\,{\small \cdot }\sum_{j=0}^{\mathcal{M(}m\mathcal{)}%
-1-q}3^{2j}\,\mathbb{C}_{1/3}(m,q,j)\right] ,  \label{eq27}
\end{equation}%
with 
\begin{eqnarray*}
\mathbb{C}_{1/3}(m,q,j) &=&C_{3}^{(0)}(\mathsf{M}\mathcal{(}m\mathcal{)}%
+q+j,j)\,\Delta B_{2m,\,2\mathsf{M}\mathcal{(}m\mathcal{)}+1+2q+2j} \\
&&+3\,C_{1}^{(0)}(\mathsf{M}\mathcal{(}m\mathcal{)}+1+q+j,j)\,\Delta B_{2m,2%
\mathsf{M}\mathcal{(}m\mathcal{)}+2+2q+2j} \\
&&+9\,C_{3}^{(2)}(\mathsf{M}\mathcal{(}m\mathcal{)}+1+q+j,j)\,\Delta B_{2m,2%
\mathsf{M}\mathcal{(}m\mathcal{)}+3+2q+2j}.
\end{eqnarray*}%
and where we note that the sum in Eq. (26) contributes to $\mathcal{P}%
_{2m+1}(z)$ only if $m\geq 4.$

In a similar way, the polynomials $Q\,_{2m+1}\left( z\right) $ can after
some simplification be written as 
\[
Q_{2m+1}(z)=z^{m}+\tbinom{m}{3}(3m+1)\,z^{m-3}+z^{\epsilon (m+1)}\,{\small %
\cdot }\sum_{q=0}^{\mathcal{M(}m+1\mathcal{)-}2}\mathbf{C}%
_{2/3}(m,q)\,\left( z^{3}/9\right) ^{q}, 
\]%
where the coefficients $\mathbf{C}_{2/3}(m,q)$ are given by%
\[
\mathbf{C}_{2/3}(m,q)=\frac{3^{2\mathcal{M(}m+1\mathcal{)}}}{2^{2m}}%
[C_{2}^{(0)}(m,\mathcal{M(}m+1\mathcal{)}-q)+C_{2}^{(2)}(m,\mathcal{M(}m+1)\,%
\mathcal{-}1-q)] 
\]%
\ \ \ \ \ \ \ \ \ \ \ \ \ \ \ \ \ \ \ \ \ \ \ \ \ \ \ \ \ \ \ \ \ \ \ \ \ \
\ \ \ \ \ \ \ \ \ \ \ \ \ \ \ \ \ \ \ \ \ \ \ \ \ \ \ \ \ \ \ \ \ \ \ \ \ \
\ \ \ \ \ \ \ \ \ \ \ \ \ \ \ \ \ \ \ \ \ \ \ \ \ \ \ \ \ \ \ \ \ \ \ \ \ \
\ \ \ \ \ \ \ \ \ \ \ \ \ \ \ \ \ \ \ \ \ \ \ \ \ \ \ \ \ \ \ \ \ \ \ \ \ \
\ \ \ \ \ \ \ \ \ \ \ \ \ \ \ \ \ \ \ \ \ \ \ \ \ \ \ \ \ \ \ \ \ \ \ \ \ \
\ \ \ \ \ \ \ \ \ \ \ \ \ \ \ \ \ \ \ \ \ \ \ \ \ \ \ \ \ \ \ \ \ \ \ \ \ \
\ \ \ \ \ \ \ \ \ \ \ \ \ \ \ \ \ \ \ \ \ \ \ \ \ \ \ \ \ \ \ \ \ \ \ \ \ \
\ \ \ \ \ \ \ \ \ \ \ \ \ \ \ \ \ \ \ \ \ \ \ \ \ \ \ \ \ \ \ \ \ \ \ \ \ \
\ \ \ \ \ \ \ \ \ \ \ \ \ \ \ \ \ \ \ \ \ \ \ \ \ \ \ \ \ \ \ \ \ \ \ \ \ \
\ \ \ \ \ \ \ \ \ \ \ \ \ \ \ \ \ \ \ \ \ \ \ \ \ \ \ \ \ \ \ \ \ \ \ \ \ \
\ \ \ \ \ \ \ \ \ \ \ \ \ \ \ \ \ \ \ \ \ \ \ \ \ \ \ \ \ \ \ \ \ \ \ \ \ \
\ \ \ \ \ \ \ \ \ \ \ \ \ \ \ \ \ \ \ \ \ \ \ \ \ \ \ \ \ \ \ \ \ \ \ \ \ \
\ \ \ \ \ \ \ \ \ \ \ \ \ \ \ \ \ \ \ \ \ \ \ \ \ \ \ \ \ \ \ \ \ \ \ \ \ \
\ \ \ \ \ \ \ \ \ \ \ \ \ \ \ \ \ \ \ \ \ \ \ \ \ \ \ \ \ \ \ \ \ \ \ \ \ \
\ \ \ \ \ \ \ \ \ \ \ \ \ \ \ \ \ \ \ \ \ \ \ \ \ \ \ \ \ \ \ \ \ \ \ \ \ \
\ \ \ \ \ \ \ \ \ \ \ \ \ \ \ \ \ \ \ \ \ \ \ \ \ \ \ \ \ \ \ \ \ \ \ \ \ \
\ \ \ \ \ \ \ \ \ \ \ \ \ \ \ \ \ \ \ \ \ \ \ \ \ \ \ \ \ \ \ \ \ \ \ \ \ \
\ \ \ \ \ \ \ \ \ \ \ \ \ \ \ \ \ \ \ \ \ \ \ \ \ \ \ \ \ \ \ \ \ \ \ \ \ \
\ \ \ \ \ \ \ \ \ \ \ \ \ \ \ \ \ \ \ \ \ \ \ \ \ \ \ \ \ \ \ \ \ \ \ \ \ \
\ \ \ \ \ \ \ \ \ \ \ \ \ \ \ \ \ \ \ \ \ \ \ \ \ \ \ \ \ \ \ \ \ \ \ \ \ \
\ \ \ \ \ \ \ \ \ \ \ \ \ \ \ \ \ \ \ \ \ \ \ \ \ \ \ \ \ \ \ \ \ \ \ \ \ \
\ \ \ \ \ \ \ \ \ \ \ \ \ \ \ \ \ \ \ \ \ \ \ \ \ \ \ \ \ \ \ \ \ \ \ \ \ \
\ \ \ \ \ \ \ \ \ \ \ \ \ \ \ \ \ \ \ \ \ \ \ \ \ \ \ \ \ \ \ \ \ \ \ \ \ \
\ \ \ \ \ \ \ \ \ \ \ \ \ \ \ \ \ \ \ \ \ \ \ \ \ \ \ \ \ \ \ \ \ \ \ \ \ \
\ \ \ \ \ \ \ \ \ \ \ \ \ \ \ \ \ \ \ \ \ \ \ \ \ \ \ \ \ \ \ \ \ \ \ \ \ \
\ \ \ \ \ \ \ \ \ \ \ \ \ \ \ \ \ \ \ \ \ \ \ \ \ \ \ \ \ \ \ \ \ \ \ \ \ \
\ \ \ \ \ \ \ \ \ \ \ \ \ \ \ \ \ \ \ \ \ \ \ \ \ \ \ \ \ \ \ \ \ \ \ \ \ \
\ \ \ \ \ \ \ \ \ \ \ \ \ \ \ \ \ \ \ \ \ \ \ \ \ \ \ \ \ \ \ \ \ \ \ \ \ \
\ \ \ \ \ \ \ \ \ \ \ \ \ \ \ \ \ \ \ \ \ \ \ \ \ \ \ \ \ \ \ \ \ \ \ \ \ \
\ \ \ \ \ \ \ \ \ \ \ \ \ \ \ \ \ \ 
\[
+\frac{6^{2q}}{2^{2\mathcal{M(}m+1\mathcal{)}}}\,[\Delta B_{2m,\,2\,\mathsf{M%
}(m+1)+2q}-\left\{ 2+3{\small \mathsf{M}(m+1)+}3q\right\} \,\Delta B_{2m,2\,%
\mathsf{M\,}(m+1)+1+2q}] 
\]

\[
+\frac{3^{2q+2}}{2^{2m}}{\small \cdot }\sum_{j=0}^{\mathcal{M(}m\mathcal{)-}%
2-q}3^{2j}\,\mathbb{C}_{2/3}(m,q,j), 
\]%
with\ 

\[
\begin{array}{c}
\mathbb{C}_{2/3}(m,q,j)= \\ 
=\left\{ {\small C}_{2}^{(0)}{\small (j+\,\mathsf{M}(m+1)+1+q,\,j+1)+C}%
_{2}^{(2)}{\small (j+\,\mathsf{M}(m+1)+1+q,\,j)}\right\} {\small \Delta B}%
_{2m,\,\,2\mathsf{M}(m+1)+2+2q+2j} \\ 
\!-3\left\{ {\small C}_{4}^{(0)}{\small (j+\,\mathsf{M}(m+1)+1+q,\,j+1)+C}%
_{4}^{(2)}{\small (j+\,\mathsf{M}(m+1)+1+q,\,j)}\right\} {\small \Delta B}%
_{2m,\,\,2\mathsf{M}(m+1)+3+2q+2j},%
\end{array}%
\]%
and where we note that the sum over $\mathbf{C}_{2/3}(m,q)$ is empty for $%
m\leq 5$ and the relation%
\[
\tfrac{1}{{\small 4}}[\Delta B_{2m,\,2m\,-2}-(3m-1)\Delta B_{2m,\,2m\,-1}+%
\tfrac{1}{2}m\,(27m-7)]=\tbinom{m}{3}(3m+1), 
\]%
has been used to simplify the coefficient of $z^{m-3}$ in the expression for 
$Q_{2m+1}(z)$. \ 

It does not appear to be possible to further simplify either of the $\mathbf{%
C}_{1/3}(m,q),\ \mathbf{C}_{2/3}(m,q)$ coefficients. \ In practical terms,
the expressions for $\mathcal{P}_{2m+1}(z)$ and $Q_{2m+1}(z)$ are seen to
exhibit a structure which depends on the discontinuous functions $\mathsf{M}%
(m),$ $\mathcal{M(}m\mathcal{)}$ and $\epsilon (m)$ and as a result these
functions are in practical terms responsible for the structural nature of
those polynomial sequences.\ \ Ultimately however, the underlying structure
and properties of these polynomials rests upon the nature of the Airy
function itself.

\appendix

\begin{center}
Appendix 1\bigskip
\end{center}

Having obtained general expressions for the polynomials of odd order i.e. $%
\mathcal{P}_{2m+1}(z)$ and $Q\,_{2m+1}\left( z\right) $ , the even order
polynomials can be obtained from the recurrence relations given above. \ We
have%
\[
\mathcal{P}_{2m}(z)=\left[ \mathcal{P}_{2m+3}(z)-z\,\mathcal{P}_{2m+1}(z)%
\right] /(2m+1), 
\]%
\[
Q\,_{2m}\left( z\right) =\left[ Q\,_{2m+3}\left( z\right)
-z\,Q\,_{2m+1}\left( z\right) \right] /(2m+1). 
\]%
Alternately the even order polynomials can also be obtained using the
differential difference equations also shown above i.e. 
\[
\mathcal{P}_{2m}(z)=z\,Q\,_{2m-1}\left( z\right) +\frac{d\,\mathcal{P}%
_{2m-1}(z)}{dz}, 
\]%
\[
Q\,_{2m}\left( z\right) =\mathcal{P}_{2m-1}(z)+\frac{dQ\,_{2m-1}\left(
z\right) }{dz}. 
\]

\begin{center}
Appendix 2\bigskip
\end{center}

Expressions for the coefficients $C_{i}^{(0)}(m,p)$ and $C_{i}^{(2)}(m,p)$
have been given below. \ 

\[
C_{1}^{(0)}(m,p)=\sum_{i=0}^{m-1-p}\left[ \tbinom{2m}{i}\tbinom{m+p-i}{2p+1}%
\tfrac{\Gamma (5/3-m+p+i)}{\Gamma (2/3-m-p+i)}-\tbinom{2m}{m+p+1+i}\tbinom{%
i+2p+1}{2p+1}\tfrac{\Gamma (8/3+2p+i)}{\Gamma (5/3+i)}\right] 
\]%
\[
C_{2}^{(0)}(m,p)=\sum_{i=0}^{m-1-p}{\LARGE [}\tbinom{2m}{i}\tbinom{m-1+p-i}{%
2p}\tfrac{\Gamma (2/3-m+p+i)}{\Gamma (2/3-m-p+i)}+\tbinom{2m}{m+p+1+i}%
\tbinom{i+2p}{2p}\tfrac{\Gamma (8/3+2p+i)}{\Gamma (8/3+i)}{\LARGE ]} 
\]%
\[
C_{2}^{(2)}(m,p)=\tfrac{2}{3}\sum_{i=0}^{m-1-p}\tbinom{2m}{m+p+1+i}\tbinom{%
i+2p+1}{2p+1}\tfrac{\Gamma (8/3+2p+i)}{\Gamma (5/3+i)} 
\]%
\[
C_{3}^{(0)}(m,p)=\sum_{i=0}^{m-1-p}\tbinom{2m+1}{i}\tbinom{m+p-1-i}{2p}%
\tfrac{\Gamma (4/3+m+p-i)}{\Gamma (4/3+m-p-i)} 
\]%
\[
C_{3}^{(2)}(m,p)=\sum_{i=0}^{m-1-p}{\LARGE \{}\tfrac{1}{3}\tbinom{2m+1}{i}%
\tbinom{m+p-i}{2p+1}\tfrac{\Gamma (4/3+m+p-i)}{\Gamma (1/3+m-p-i)}+\tbinom{%
2m+1}{i+m+p+2}\tbinom{i+2p+2}{2p+2}\tfrac{\Gamma (11/3+2p+i)}{\Gamma (5/3+i)}%
{\Large \}} 
\]

\[
C_{4}^{(0)}(m,p)=\sum_{i=0}^{m-1-p}\tbinom{2m+1}{i}\tbinom{m+p-i}{2p+1}%
\tfrac{\Gamma (4/3+m+p-i)}{\Gamma (1/3+m-p-i)} 
\]%
\[
C_{4}^{(2)}(m,p)=\sum_{i=0}^{m-1-p}\tbinom{2m+1}{i+m+p+2}\tbinom{i+2p+3}{%
2p+3}\tfrac{\Gamma (11/3+2p+i)}{\Gamma (2/3+i)} 
\]%
The finite sums which make up these coefficients can also be represented by
combinations of the hypergeometric functions $_{3}F_{2}$ and $_{4}F_{3\text{ 
}}$although there appears to be little or no advantage gained in expressing
them in that way. \ In less general terms each of these sums can be written
as weighted polynomials in $m.$\ \ For example $C_{1}^{(0)}(m,p)=2^{2m}%
\tbinom{m}{m-p-1}{\small \cdot }\wp ^{(p)}(m)$ where the symbols $\wp
^{(\omega )}(m)$ or $\wp ^{\prime \,(\omega )}(m)$ indicates a polynomial in 
$m$ of order $\omega $. \ In the case of the other coefficients one finds 
\begin{eqnarray*}
C_{2}^{(i)}(m,p) &=&2^{2m}\wp ^{(2p+1)}(m)+\wp ^{\prime \,(2p+1)}(m){\small %
\cdot }\tbinom{2m}{m}, \\
C_{3}^{(i)}(m,p) &=&2^{2m}\wp ^{(2p)}(m)+\wp ^{\prime \,(2p)}(m){\small %
\cdot }\tbinom{2m+2}{m+1}, \\
C_{4}^{(i)}(m,p) &=&2^{2m}\wp ^{(2p+1)}(m)+\wp ^{\prime \,(2p+1)}(m){\small %
\cdot }\tbinom{2m+2}{m+1}.
\end{eqnarray*}


\begin{thebibliography}{9}
\bibitem{V} O. Vallee and M. Soares, \textit{Airy Functions and Applications
to Physics}, Imperial college Press (2010). \ Also see the Airy function
bibliography \
\url{http://math.fullerton.edu/mathews/n2003/airyfunction/AiryFunctionBib/Links/AiryFunctionBib\_lnk\_2.html}

\bibitem{A} E. Abramochkin, private communication.

\bibitem{AandS} M. Abramowitz and I. A. Stegun, \textit{Handbook of
Mathematical Functions}, National Bureau of Standards (1972), p. 447.

\bibitem{Faa} E. W. Weisstein, ``Fa\`{a} di Bruno's Formula.'' From
MathWorld--A Wolfram Web Resource.
\url{http://mathworld.wolfram.com/FaadiBrunosFormula.html}

\bibitem{Bell} E. T. Bell, ``Partition Polynomials'', Annals of Mathematics 
\textbf{29} (1/4), 1927--1928, pps. 38--46. \ Also see
http://en.wikipedia.org/wiki/Bell\_polynomial.

\bibitem{C} D. Cvijovic, New identities for the partial Bell polynomials,
Appl. Math. Lett. \textbf{24}, No. 9, pps. 1544-1547 (2010).

\bibitem{Bess} G. N. Watson, \textit{Theory of Bessel Functions}, Cambridge
University Press, Chap. 3, p. 79, (1966). \ Also see
http://functions.wolfram.com/03.04.20.0014.02 and
http://functions.wolfram.com/03.04.17.0005.01

\bibitem{Heavi} In this paper we are using the definition of the Heaviside
function contained in the web page at
http://en.wikipedia.org/wiki/Heaviside\_function

\bibitem{Ceil} R. L. Graham and D. E. Knuth, \textit{Concrete Mathematics},
Addison-Wesley, Reading Ma., Chap.13, (1994). \ Also see
http://functions.wolfram.com/04.02.02.0001.01Mathematica and
http://en.wikipedia.org/wiki/Floor\_and\_ceiling\_functions\#cite\_note-0.
\end{thebibliography}
\end{document}